%% file: WVR_PASA.tex
\documentclass[a4paper,twoside]{article}
\newcommand{\affil}[1]{$^{\rm #1}$}
\usepackage{multirow}
\usepackage{tabularx}
\usepackage{booktabs} 

\makeatletter
\newcommand\textsubscript[1]{\@textsubscript{\selectfont#1}}
\def\@textsubscript#1{{\m@th\ensuremath{_{\mbox{\fontsize\sf@size\z@#1}}}}}
\newcommand\textbothscript[2]{%
  \@textbothscript{\selectfont#1}{\selectfont#2}}

\def\@super@sub#1_#2{\textbothscript{#1}{#2}}
\def\@sub@super#1^#2{\textbothscript{#2}{#1}}
\makeatother  

\usepackage[authoryear]{natbib}
\bibpunct{(}{)}{;}{a}{}{,}

\usepackage{graphicx}

\usepackage{amsmath}  
\usepackage[mathscr]{eucal}
\date{} 
\let\oldsqrt\sqrt
\def\sqrt{\mathpalette\DHLhksqrt}
\def\DHLhksqrt#1#2{%
\setbox0=\hbox{$#1\oldsqrt{#2\,}$}\dimen0=\ht0
\advance\dimen0-0.2\ht0
\setbox2=\hbox{\vrule height\ht0 depth -\dimen0}%
{\box0\lower0.4pt\box2}}

\title{\large\bf\flushleft Water Vapour Radiometers for the Australia Telescope Compact Array}
\author{\parbox{\textwidth}{\flushleft
\vspace{-0.5cm}
{\it Balthasar T. Indermuehle\affil{A,B}, Michael G. Burton\affil{B}, and Jonathan Crofts\affil{C}}\\
\vspace{0.4cm}
{\small \affil{A}\,CSIRO Astronomy and Space Science CASS, Epping, Email: balt.indermuehle@csiro.au}\\
{\small \affil{B}\,School of Physics, University of New South Wales, Sydney}\\
{\small \affil{C}\,ASTROWAVE Pty Ltd}}}
\begin{document}
%
\twocolumn[
\begin{changemargin}{.8cm}{.5cm}
\begin{minipage}{.9\textwidth}
\vspace{-1cm}
\maketitle
%
%
\small{\bf Abstract:}
We have developed Water Vapour Radiometers (WVRs) for the Australia Telescope Compact Array (ATCA) that are capable of determining signal path length fluctuations by virtue of measuring small temperature fluctuations in the atmosphere using the 22.2 GHz water vapour line for each of the six antennae. By measuring the line of sight variations of the water vapour, the induced path excess and thus the phase delay can be estimated and corrections can then be applied during data reduction. This reduces decorrelation of the source signal. We demonstrate how this recovers the telescope's efficiency as well as how this improves the telescope's ability to use longer baselines at higher frequencies, thereby resulting in higher spatial resolution. A description of the WVR hardware design, their calibration and water vapour retrieval mechanism is given.


\medskip
\medskip
\end{minipage}
\end{changemargin}
]
\small

\section{Introduction}

\subsection{Overview}
The Australia Telescope Compact Array (ATCA) is a radio interferometer capable of observing in the centimetre and millimetre wavelength regime of the electromagnetic spectrum. It was for a little over a decade the only millimetre interferometer in the southern hemisphere.
\\
Variations in the water vapour content in the atmosphere along the lines of sight from the source being observed to each antenna lead to fluctuations in the refractive index, and so to path delays and phase fluctuations. At millimetre wavelengths, these fluctuations are rapid, requiring frequent measurement of a calibrator source in order to be followed. This may be as often as every 5-10 minutes, depending on weather conditions and frequency. This can consume up to a quarter of the observing time. The ATCA has 3 mm receivers (tuneable from 83 to 105 GHz), 7 mm receivers (tuneable from 30 to 50 GHz) and 15 mm receivers (tuneable from 16 to 26 GHz). They are mounted on a turret which moves the feed for the requested receiver on axis automatically. The water vapour induced phase variations increase with the length of the baseline between antennae.  With the Australia Telescope Compact Array, the largest useable baseline at 3 mm is found to be $\sim$ 300m, yet the maximum baseline separation of the five antennae equipped with 3 mm receivers is 3km (twice that, 6 km, for 7 mm and 15 mm), and this is used regularly at centimetre wavebands. If the maximum available baselines could be used at millimetre wavelengths it would result in up to a factor of ten improvement in the spatial resolution for millimetre imaging at 3 mm.  On shorter baselines improved phase stability would also improve the signal to noise ratio of millimetre-wave observations.\\
\\
The application of phase corrections derived from WVR data may also extend the millimetre wave observing, allowing the telescope to be used in a wider range of conditions. In 2006, 33\% of proposals in the April semester involved observations at 3mm: in 2007 the figure was 25\%. In 2008, 37\% of Australia Telescope Compact Array proposals requested 3mm observing, and fully 45\% requested 3mm and/or 7mm observations. These figures are comparable throughout the years up until the April 2011 semester. The interest in millimetre observing has remained strong with the installation and commissioning of the Compact Array Broadband Backend (CABB) in 2009 (\cite{2011MNRAS.416..832W}).\\

\subsection{The Atmosphere}
The various constituents of the atmosphere all have an effect on the transmission of signals from space. In the millimetre region (between about 16 GHz to 300 GHz), the main contributors to atmospheric attenuation are O\textsubscript{2} and H\textsubscript{2}O. For $\lambda \leq$ 1.5 cm, the optical depth of the troposphere becomes significant, leading to increased system temperatures due to atmospheric emission (\cite{Carilli:1999fk}). The refractivity of water vapour at these frequencies is about 20 times greater than in the near-infrared or optical regimes (\cite{Thompson:2001fk}). Figure \ref{figure1} shows the sky brightness caused by these opacity contributions.  The water vapour lines at 22.2 and 183.3 GHz are prominent.\\
\\
When evaluating the atmospheric conditions, several models were used to characterise the temperature differences arising due to variations in the water vapour content along the lines of sight.  The first was the Millimeter Propagation Model (MPM) developed by \cite{Liebe:1985fk} and implemented in the Miriad software package. While accurate enough for purposes of filter optimisation, the model has been improved upon by \cite{Pardo:2001fk} who developed the Atmospheric Transmission at Microwaves (ATM) model used to simulate PWV induced excess temperatures. The well characterised VanVleck -- Weisskopf (VVW) profiles were also used to determine the model atmosphere parameters for the water vapour calibration factors as well as for the weighting coefficients determined in \S \ref{coefficients} to combine measurements made at several frequencies across the 22.2 GHz water line. The same VVW profiles used for the coefficients were also incorporated into the ATM model.

\section{WVR Design}
\subsection{Design principle}
The WVRs developed for ATCA are an uncooled design based on four filters placed between the frequencies of 16 -- 26 GHz, each with 1 GHz of bandwidth to increase the sensitivity. This allows the WVRs to measure small variations in the spectrum of the 22.2 GHz water vapour emission and derive phase variations based on these measurements. They are co-located with the science receivers, with the WVR feeds offset by 13 cm to the on-axis millimetre receiver being used, and hence look through almost the same beam onto the sky. They operate at frequencies the antennae were designed to perform at.  They further are low maintenance due to the lack of a cryogenically cooled stage.  By using a differential water vapour determination, absolute measurements are not required. We use them to measure the minute temperature differences caused by the fluctuations in the water vapour between the lines of sight for each antenna pair.

\subsection{Determining a figure of merit}
\label{requiredprecision}
We introduce a figure of merit to model performance of the WVRs in order to determine the accuracy to which the temperature fluctuations must be measured. The expected level of decorrelation of the incoming signal due to atmospheric effects can be expressed using the Ruze formula (\cite{Ruze:1966kx}) shown in equation \ref{Ruze}.  This equation states that for a signal reflected off an antenna surface of area $A_{0}$, the aperture efficiency is $A/A_0$.  

\begin{equation}\label{Ruze}
A = A_0 e^{-(\frac{4\pi\sigma}{\lambda})^2}
\end{equation}

where $\sigma$ is the RMS surface accuracy of the antenna. Applying this to interferometry, the phase shift inducing error term which for single dish application is $4\pi\sigma/\lambda$ is halved to $2\pi\sigma/\lambda$ because the locus of the error is above the antenna and the path error is thus induced only once. $\sigma$ is now the path difference caused by the water vapour fluctuations.  By expressing the required precision N as a fraction of the wavelength $\lambda$, i.e. $1/N = \sigma/\lambda$, the figure of merit is obtained, hereafter referred to as the correlation efficiency $\epsilon$:

\begin{equation}\label{epsilon}
\epsilon = e^{-(\frac{2\pi}{N})^2}
\end{equation}

Equation \ref{epsilon} is plotted in Figure \ref{fig:epsilon} to show the $\lambda$ fraction to which one needs to correct in order for the WVRs to provide useful levels of improvement: For a correlation efficiency of at least $\epsilon=0.5$, the water vapour induced path delay must be determined to better than $N_{threshold}=\lambda/7.5$, to achieve improvement to $\epsilon=0.67$, $N_{threshold}=\lambda/10$ must be achieved and to obtain a virtually perfect correlation efficiency of $\epsilon=0.9$, corrections to $N_{threshold}=\lambda/20$ must be measured.

From this we can determine the required sensitivity limits by evaluating the temperature fluctuations that the WVRs need to be able to measure in order to yield a correlation efficiency of at least $\epsilon=0.5$. This result is shown in Figure \ref{deltaPWV}, where we show the temperature excess for several models with differing PWV and pressure (as shown in the caption) at an observing wavelength of 3 mm for three different values of path differences. These values correspond to fractions of the observing wavelength of 1/7.5, 1/10 and 1/20 (equal to the three thresholds above). The relevant PWV excesses have been determined using Equation \ref{Lvxsimple}, which as we discuss in \S\ref{sec:3} relates the path length difference to the excess water vapour column.

\subsection{Filter optimisation}
The original filter locations were based on a design by Robert J. Sault. The frequencies were chosen to cover the water line in its most sensitive regions based on atmospheric modelling and a 10 year archive of radio sonde data providing the input for the model. We have repeated these optimisation steps but modified them for a wider filter bandwidth with 3 filters placed across the line and one filter on the continuum. We have also identified a number of RFI sources (chiefly communication satellites) and have managed to keep the filter locations outside the regions of the commercially allocated radio spectrum. The filter placement is illustrated in Figure  \ref{fig:filterplacement}. Shown here are the overall water line excess variations for a 10 year period using radio sonde data as input for the model. The dotted lines represent the maxima/minima values encountered in the entire data set.

\subsection{Determining Noise Limits}
\label{sec:sensitivity}
The theoretical sensitivity of an ideal total-power radiometer with no gain fluctuations is calculated as (\cite{Ulaby:1981fk}):

\begin{equation}\label{deltaT}
\Delta T = \frac{T_a + T_e}{\sqrt{\Delta f \tau}}
\end{equation}

where $\Delta T$ is the minimum detectable change in the radiometric antenna temperature $T_a$, $T_e$ is the noise temperature of the entire signal chain, $\tau$ is the integration time and $\Delta f$ is the bandwidth of the system. This is shown in Figure \ref{fig:sensitivity}.

$T_e$, the total noise temperature, is calculated as a cascaded system taking each component's noise temperature contribution in the signal chain into account as given by the Friis formula (\cite{Kraus:1966fk}):

\begin{equation}\label{Te}
T_e = T_{e_1} + \sum_1^N \frac{T_{e_{(N+1)}}}{\prod_1^NG_N}
\end{equation}

where $T_{e_1}$ is the noise temperature of the first stage, $T_{e_N}$ and $G_N$ are the noise temperature and the gain of the $N^{th}$ stage respectively.\\
\\
To specify the noise limitation, the fundamental equation determining a radio receiver noise floor is applicable.  The power received is determined by:

\begin{equation}\label{rxpower}
P=kT \Delta f
\end{equation}

where $k$ is the Boltzmann constant, $T$ is the temperature in K and $\Delta f$ is the bandwidth of the receiver in Hz.  Note that in high frequency engineering, it is useful to express the power and noise ratios as referenced to 1 milliwatt (abbreviated as dBm).  $P$ hence is converted to mW for all calculations.  Assuming 290 K for the temperature and 1 GHz of bandwidth, the minimum equivalent input noise of the water vapour radiometer is --83 dBm, also noted on the receiver diagram in Figure \ref{fig:schematics}.  The noise floor $N_f$ can then be determined:

\begin{equation}\label{noisefloor}
N_f = -174\, \mathrm{dBm}   +   N_{fig} +   10 \log_{10}(\Delta f) \, \mathrm{[dBm]}
\end{equation}

where $N_{fig}$ is the noise figure of the system and --174 dBm corresponds to the input noise of 1 Hz of bandwidth at room temperature (290 K). The noise figure, noise floor and receiver temperatures have been determined for all channels and units. The values for Unit 1 are listed in Table \ref{tab:noisefigures}. Please refer to \cite{Indermuehle:2011nf} for a listing of all units.

\subsection{Hardware}\label{sec:hardware}
The uncooled radiometers are designed to reside inside the vertex room, attached to the millimetre receiver system mounting cage, also referred to as the ``millimetre package''.  The WVR feed horn is situated in such a way that regardless of which mm receiver is on axis, the WVR feed is offset by no more than 12 cm from the axis.  According to \cite{Cooper:1992bh}, the ATCA antennae have a diameter of 22 m, and an $f$ ratio of 2.0, so the beams for the receiver and WVR should be separated only by 5.4 m at 2,000 m altitude, resulting in a theoretical angular offset of $9.3'$. The geometrical column of the WVR feed should thus be overlapping that of the science receivers by 68\%. The majority of troposphere induced path delays originate from below 2,000 m altitude. Hence the WVRs are measuring fluctuations through virtually the same column of atmosphere as the science receivers. The beams are completely separated above an elevation of $\sim$\,8,000 m. We have experimentally verified the feed positions in February 2013 using geostationary satellites and determined that the feeds are offset by $7.5'$ from the science receivers. The overlap thus is better than calculated and is offset only by 4.3 m at 2,000 m altitude, providing geometrical footprint overlap of 73\%. \\
\\
To illustrate the signal path and the components involved in the water vapour radiometers, as well as to visualise the insulation characteristics of the package, we now walk through the signal path while simultaneously looking at the following two figures: 
\begin{itemize}
\item The radiometer schematics in Figure \ref{fig:schematics}, starting at the top, 
\item Unit 4 in Figure \ref{fig:unit4photo}.
\item Refer to Figure \ref{fig:section1} for a cut-away render through a WVR with its insulation layers.
\end{itemize}

The signal path is entered at the feed horn (only shown on the schematic) where a weak signal of --83 dBm at 1 GHz of bandwidth (corresponding to the widest filters in the system) is received.  The waveguide (numbered 1)  feeds this signal into the first two low noise Miteq amplifiers (LNA's, numbered 2) which amplify the signal each by +36 dB.  The signal continues via an Inmet matching pad through a 26 GHz low pass filter (numbered 3) and another matching pad, introducing a total loss of 25 dB.  Then the first MCLI splitter is reached (numbered 4) which splits off the total power signal from the other filters. The total power channel now directly feeds into the Herotek tunnel diode detector and from there into the Analog Devices ultra precision operational amplifiers from where the signal is digitised and fed into the dataset (this last step is not shown). A dataset refers to a piece of hardware which converts the digital signals received from various systems in the antenna into a databus protocol which can be ingested by the centralised monitoring system at the ATCA (MoniCA). Following the other signal path after the total power splitter, another matching pad is entered and another Miteq LNA (numbered 5) amplifies the signal by +36 dB, then the signal is fed through a Ditom isolator, an MCLI four-way splitter (numbered 6) and into the ASTROWAVE band pass filters (numbered 7), one for each of the filter frequencies of 16.5, 18.9, 22.9 and 25.5 GHz respectively.  Then the signal for each filter follows the Herotek tunnel diode detectors into the amplifiers and then into the datasets where the data are obtained via the millimeter receiver package data bus.  The arrangement of the WVR feed as well as the hardware locations are shown in Figure \ref{fig:topdewar}, showing a view from the top of the focal plane into the feed horns situated on top of the millimeter package dewar, as well as in Figure \ref{fig:sideview} the WVR unit can be seen mounted on the side of the millimeter receiver package.

The required temperature stability was derived from the overall gain stability requirements discussed by \cite{Bremer:1997fk}. Considering receiver noise and integration time, they suggest a conservative fractional stability of $10^{-4}$ as the maximum allowed fractional change in gain between calibrations. By far the most temperature sensitive components of the receiver are the Miteq JS microwave amplifiers with a quoted gain shift of 0.04 dB/K. Having tested all other components in the chain this represents the defining system gain stability with respect to temperature. This translates into 1 \% gain change for a shift of 1 K.  Therefore, to achieve no more than $10^{-4}$ gain shift, the temperature must be stabilised to about 10 mK at the Miteq JS amplifier position on the RF plate. This temperature variation mainly comes from changing thermal gradients along the RF plate which add to any temperature variation at the control point so a conservative figure of 1 mK stability was chosen for the control point on the RF plate.

These limits were verified by calculating the Allan variance and deviation shown in equations \ref{allanvariance} and \ref{allandeviation}. This is a useful metric because similarly to a power spectrum it highlights behaviour in the time domain such as externally caused temperature oscillations e.g.  by the vertex room air conditioning system.  The Allan deviation is calculated as the square root of the Allan variance, which in turn is one half the time average of the squares of the differences between successive data points of the temperature fluctuations, sampled over the sampling period as shown in equation \ref{allanvariance}.

\begin{equation}\label{allanvariance}
\sigma_T^2(\tau) = \frac{1}{2} \langle (\overline{T_{N+1}} - \overline{T_N})^2 \rangle
\end{equation}

where $\overline{T_N}$ is the $N^{th}$ fractional temperature average over the observation time $\tau$. From this follows the Allan deviation:

\begin{equation}\label{allandeviation}
\sigma_T(\tau) = \sqrt{\sigma_T^2(\tau)}
\end{equation}

The RF Plate temperature control point data was found to be well within the required limits at values between 0.01 and 0.4 mK. One example plot is shown in Figure \ref{fig:allen_u1}.

The environmental conditions taken into account for the thermal stabilisation design was assuming an ambient temperature range of 10 K.  This has been verified from temperature monitoring data of the vertex rooms. The cycling of the air conditioners also mean that these temperature variations can occur on a short time scale compared to the calibration rate of the WVRs so they need to be able to hold 10 mK at the Miteq JS amplifiers with an external temperature change of this magnitude between calibrations. The housing insulation and intermediate temperature controlled box were designed to achieve this. The ideal shell material is anisotropic with good insulation `across' the medium and good conductivity `along' the medium (i.e. around the RF plate) to minimize temperature gradients.  A useful approximation to this is several alternate layers of thermal insulator and conductor.  This presents a challenge if easy access to the RF plate is to be maintained.  A practical solution was to build two insulating layers and three conductive (aluminium) layers with the inner two actively temperature controlled.  The minimum temperature gradient change was successfully achieved using this method. Refer back to Figure \ref{fig:section1}: The top halves of the conducting layers were rendered as a transparent plastic in this visualisation to illustrate the layering.  The hollow spaces in between are foam filled in the real units.\\
\\
For both gain stability and temperature stability the +10V drive voltage to the Miteq JS amplifiers needs to be exceedingly stable, because changes in power dissipation of RF components would change local temperature.  Great care was taken to stabilise this using precision voltage references installed on the temperature stabilized RF plate printed circuit board (PCB).  This helps to maintain the temperature stabilised environment with all of the control semi-conductors (which all have varying power dissipation) outside of the thermal enclosure and placed directly on the main heat sink.\\
\\
To thermally isolate the RF plate from the environment, all electrical conductors leading into the RF electronics were fabricated as thin traces on a 200 mm long specially manufactured flexible Kapton printed circuit.  The input coaxial feed is fabricated from stainless steel for low thermal conductivity, with thin silver layers on the centre conductor and inside of the outer conductor for low electrical loss.  This  reduces thermal leakage but it is still the weakest link in the thermal design because this cable penetrates the insulating layers.\\
\\
Sharp cut off filters to minimise overlap between the two closest channels were needed.  A 7th order Chebychev filter design was chosen as they give the sharpest cut off for a given order with, theoretically, no out of band responses and the choice of 1 dB ripple in the pass band is dampened out by the insertion loss.  7th order Micro-Strip design also gives a rather high insertion loss at this frequency (5 to 7 dB) but this is at high signal level at the end of the signal chain and there is sufficient gain available to render the loss minor.\\
\\
The filters were implemented as Micro-Strip filters etched on 0.25 inch backed `Duroid' (i.e.  ceramic loaded PTFE\footnote{PTFE = Polytetrafluoroethylene, more commonly known as the brand name Teflon.}) substrate.  This gave a small physical size, helping keep the overall WVR package small.  Also, by keeping the RF plate as small as possible, the temperature gradients induced by a given thermal leakage remain smaller.  Further, this makes for a mechanically very solid, and stable, filter.\\
\\
There was some concern regarding thermal stability of this substrate material but bench tests of a completed filter over a 50 K range indicated that the insertion loss had a temperature coefficient of only 0.01 dB /  K, or 0.2\%, which is a factor of 5 better than the RF amplifier's temperature coefficient of 0.04 dB / K (which amounts to about 1\%).

\subsection{Calibration}

Calibration of the water vapour radiometers was achieved by measuring the individual channel voltage at two known set points.  The first set point measures the hot load (or paddle), consisting of activating the paddle mechanism and reading the temperature of a probe inside the absorber foam that is mounted on the paddle.  There is also a case temperature probe (called T5) on each water vapour radiometer which can be used to indicate the ambient temperature (and thus approximates the hot load temperature).  The colder set point is achieved by setting a styrofoam box on top of the feed horn with a microwave absorber inside submerged in a sufficient amount of liquid nitrogen so that the absorber along with the LN\textsubscript{2} becomes optically thick.\\
\\
With cold and hot load on the feed, the spillover contribution is not a factor and the gain $G$ can be determined via the Y factor:

\begin{equation}\label{Yfactor}
\frac{T_{Hot} + T_{Rec}}{T_{Cold} + T_{Rec}} = \frac{V_{Hot}}{V_{Cold}} = Y
\end{equation}

where $T_{Hot}$ is the hot load temperature, $T_{Cold}$ is the cold load temperature and

\begin{equation}\label{TRec}
T_{Rec} = \frac{T_{Hot} - Y T_{Cold}}{Y-1}
\end{equation}

$V_{Hot}$ and $V_{Cold}$ are the hot and cold load voltages measured and 

\begin{equation}\label{Gain}
G = \frac{T_{Hot} + T_{Rec}}{V_{Hot}}
\end{equation}

The gains and Y factors have been determined three times, the data for Unit 7 (antenna 4) and its first two calibrations are shown in Table \ref{tab:gaincals}. The Y factor is seen to change by less than 1\% between the two calibrations, which were spaced 3 months apart. A third calibration was done 7 months after the first calibration and found no significant change in Y factors.  This means that with the hot load alone, gain drift corrections can be applied. This, as well as a table with the full calibration data for all Units, is discussed in further detail in \cite{Indermuehle:2011nf}.\\
\\
After each calibration series, a zenith sky spectrum was obtained for each antenna.  Figure \ref{unit1_zenith_spectrum} shows this spectrum for Unit 1.  The sky temperatures in the channels follow the model atmosphere as expected and outline the shape of the H\textsubscript{2}O line on top of the $\nu^2$ continuum.  The sky temperature in the total power channel (spanning the full 16 -- 26 GHz) is at a similar level to the highest continuum point at 25.5 GHz, and the shape of the water line is traced in all filter positions.  A further representation of these data for all Units is shown in Figure \ref{fig:spectrum}, which also shows the ATM derived model atmospheres for 5~mm, 30~mm and 23~mm of PWV reflecting the site minimum PWV, the maximum PWV beyond which millimetre observations are discontinued as well as the conditions when the measurements were taken, respectively.

\subsection{Sky Dips}
During the calibration measurements several sky dips were obtained, one set of which is depicted in Figure \ref{fig:skydip} for each of the WVR units.  These are measurements of the sky flux obtained at a number of zenith angles, from overhead to 13 degrees above the horizon.  Based on the skydip data, each channel for each WVR unit had its optical depth ($\tau$) and spillover term ($T_{S}$) determined by a least squares fit of equation \ref{spillover} to the data;

\begin{equation}\label{spillover}
T_{sec(\theta)} = T_{S} + T_{A}(1-e^{-\tau sec \theta })
\end{equation}

which, for small $\tau$ becomes

\begin{equation}\label{spillover2}
T_{sec(\theta)} = T_{S} + T_{A}(\tau sec\theta).
\end{equation}

Here $T_{A}$ is the atmospheric temperature (assumed constant in the emitting layer) and $\theta$ is the zenith angle.

The slope of the curves in Fig.~\ref{fig:skydip}  yields the optical depth and the intercept with the $x$-axis the spillover temperature.  The results from the fits, including the errors, are also listed in Table \ref{tab:spillover}. The optical depths determined are further illustrated in Figure \ref{fig:tau}, where they are compared to optical depths calculated for three model atmospheres.  The measured optical depths are seen to be, in general, well represented by the models for between 28\,mm and 39\,mm of PWV\@.  However, it is also clear that a precise determination of PWV (and hence $\tau$) cannot be obtained from such a comparison.   In other words, the WVR data cannot be used for an absolute determination of the optical depth, and hence in the path lengths traversed by the radiation through the atmosphere. Only differential path lengths, obtained by subtracting the offsets between each pair of antennae, can be accurately derived, as we demonstrate in \S\ref{dcoffset}.

The spillover temperatures are also listed in Table \ref{tab:spillover}.  They are seen to be negative, and indeed most negative for the 22.9\,GHz filter, in the water vapour line itself.  The magnitudes of the derived $T_S$ are, however, only a small fraction (between 1\% and 7\%) of the ground temperature of $\sim$290\,K\@.  That the values are negative is likely the result of the microwave absorber for the cold load (submerged in liquid nitrogen) being warmer than the theoretical 77\,K of LN$_2$, with water freezing to the outside of the styrofoam box, and so presenting elevated temperatures to the radiometers.  This then results in negative values for $T_S$ when assuming 77\,K for the temperature of the cold load; the true values could be obtained by adding the error in $T_{cold}$ (if it were known). There is also variation in the WVR receiver temperatures, assumed constant in this analysis, which is of order 1 -- 3.5\% (see Table \ref{tab:gaincals}).  This further adds to the uncertainty in the determination of $T_S$.

\subsection{Time domain analysis of the filter signal}
A wavelet analysis has also been applied to the input signals received by the WVR units in order to verify that the signals they record are indeed dominated by the emission from the sky, and that they are tracking variations in the sky emission, as opposed to locally-based signals in the antenna.  Wavelet analysis provides a mean of examining the variation of power received within a time series (e.g. \citealp*{Torrence:1998fk}).  This allows determination of not only the frequencies on which the dominant power is received by a system  (as, e.g.\ a Fourier analysis would show), but also the time scales on which that power itself varies. We have used a Morlet transform for this analysis (which is equivalent to a plane wave modulated by a Gaussian) using IDL-based routines\footnote{See http://paos.colorado.edu/research/wavelets/} to process the input time series.  We show here an analysis based on the signal received through the 25.5\,GHz filter on each WVR unit, as well as, for comparison, the signal from the temperature sensor on the outer shell of each WVR unit.  The latter is tracking variations in the temperature of the receiver cabin (or vertex room).

Figure  \ref{wavelets_RF} shows the wavelet analysis for the raw voltages measured through the 25.5\,GHz filter for a 12\,hr period on 23 May 2012 for each of the 6 antennae.  The other filters show a similar response.  Time is plotted on the $x$-axis and period on the $y$-axis and the intensity of the Morlet transform indicates where the power in the input signal arises in this domain.  To first order all WVR units show a similar pattern.  A series of  weather systems passed across the observatory over the second half of the day, for instance a weather front at $\sim$04:05\,UT, and these are clearly tracked by each WVR unit, with substantial power fluctuations evident.   These short-lived events generally last for $\sim 15$ minutes and show shorter period variations (1-10 min), compared to the tens of minute timescales associated with clear sky conditions.  Antennae 2 and 3 were stowed during this time period, and so measure a signal from the zenith, while the other 4 antennae were tracking astronomical sources.  The wavelet transforms for these two antennae are similar and clearly differentiated from the other four, as would be expected if they are pointed in different directions.  Furthermore, close inspection of the times that each system passes across each antenna shows a few minute differences, running from antenna 6 to antenna 1 (a distance of 6\,km), consistent with wind speeds of 10--20\,m/s and weather systems arriving from the west.  Importantly, notably absent from the plots is any modulated power which might arise from systematic effects, such as thermally induced oscillations in the receiver cabin due to the air conditioners.  Also absent is any signal attributable to RFI\@.

In Figure \ref{wavelets_T5} is shown the same analysis, but this time for the sensor which measures the cabin temperature.  Again these plots show similar characteristics to one another, but are clearly different to the radiometer data.  The thermal variations due to the cycling of the air conditioners are now clear (and entirely absent in the radiometer signal), as are substantial differences between the behaviour of the air conditioners in individual antennae. Comparison with Fig.~\ref{wavelets_RF} confirms that the measures undertaken to thermally stabilise the WVR units have been successful.

Because of this variation in the vertex room temperature, it is not possible to determine the stability of the radiometers by measuring the hot load. We therefore assume the measurements of the RF plate temperature, where all temperature sensitive components are located, to be a valid proxy measurement for the stability of the WVRs as analysed with the Allan variance in \S\ref{sec:hardware}.


\section{Atmospheric Phase Determination}\label{sec:3}
The line strength of the water vapour emission is described in units of brightness temperature $T_{b}$ (we also use the term ``sky temperature'' interchangeably). $T_{b}$ is defined as follows:

\begin{equation}\label{Tb}
T_{b}=\frac{c^{2}}{2k\nu^{2}}I_{\nu}
\end{equation}

where $I_{\nu}$ is the specific intensity at frequency $\nu$.  In the Rayleigh-Jeans regime ($h\nu \ll kT$), the brightness temperature of a black body is equal to its physical temperature.

Below 30 GHz, absorption and emission is dominated by the weak 6\textsubscript{16}--5\textsubscript{23} transition of H\textsubscript{2}O (\cite{Liebe:1985fk}, \cite{Pardo:2001fk}). Following \cite{Carilli:1999fk}, we arrive at the path difference (or path excess) $\mathscr{L}_{VX}$ between two lines of sight:

\begin{equation}\label{Lvx}
\mathscr{L}_{VX} = 1.763 \times 10^{-6} \frac{\Delta PWV \rho_W}{T} \; \mathsf{[mm]}
\end{equation}

where $\Delta PWV$ is the difference in the water vapour columns between the two beams in mm, $\rho_W$ is the density of water in kg/m\textsuperscript{3} and $T$ the temperature in K.  We thus obtain a path length excess in mm, and from this the phase angle difference $\Delta \Phi$ in degrees, with $\lambda$ in mm:

\begin{equation}\label{dPhi}
\Delta \Phi = 360 \frac{\mathscr{L}_{VX}}{\lambda}\; \mathsf{[deg]}
\end{equation}

From equation \ref{Lvx}, we can also establish an approximate relationship under the assumption of $T$ = 292 K:

\begin{equation}\label{Lvxsimple}
\mathscr{L}_{VX} \approx 6 \cdot \mathrm{\Delta PWV} \; \mathsf{[mm]}
\end{equation}

We now discuss how to determine $\Delta PWV$ from the WVRs and so be able to calculate the phase difference.

\section{Extracting Phase}
Phase extraction is achieved by measuring the small variations in the line temperatures in each of the four filters at 16.5, 18.9, 22.9 and 25.5 GHz between each pair of antennae.  Each channel in the WVR is sampled by the dataset sequentially.  This includes all 16 signal channels including temperature and supply voltage monitoring points, in addition to the tunnel diode voltages for the filters.  The sampling process currently is implemented as follows:

\begin{itemize}
\item For all four signal channels
\begin{itemize}
\item Initial delay of 15 ms to let the input and electronic components settle.
\item 8 samples are taken every 5 ms and averaged over a total timespan of 1,175 ms for a total of 1,880 samples.
\end{itemize}
\item The remaining WVR monitoring points are sampled with 8 samples only.
\end{itemize}

Each of the important signal channels is sampled for 1,175 ms. This adds up to 4,700 ms, leaving 300 ms to sample the remainder of the channels.  Referring to the sensitivity resulting from the radiometer equation as shown in Figure \ref{fig:sensitivity}, this sampling rate yields a sensitivity of about 12.1 mK for a $T_{rec}$ of 400 K and 8.7 mK for a $T_{rec}$ of 290 K. Refer to Table \ref{tab:noisefigures} for the typical $T_{rec}$ that have been measured.

\subsection{DC Offset}
\label{dcoffset}
The measured signal includes unwanted contributions, such as the spillover and receiver terms.  They are difficult to estimate and moreover may vary with time. We work with the differences in signal and assume changes in such quantities are slow compared to the water vapour induced fluctuations so that they cancel.  We see in the next section that this indeed allows us to follow the phase variations.  After converting the raw voltages into sky temperatures using the gain calibration factors derived from the hot and cold load measurements, the total temperature $T_T$ measured in each filter can be defined as:

\begin{equation}\label{temp1}
T_T = T_{D} + T_V + T_S
\end{equation}

where $T_D$ is the dry air component, $T_V$ is the H\textsubscript{2}O line temperature and $T_S$ is the spillover temperature and other instrumental terms.  $T_D$ is assumed to be the same for each antenna looking at the same azimuth and elevation, whereas $T_S$, the spillover temperature, is antenna specific. The signals for two antennae $a1$ and $a2$ are thus

\begin{equation*}\label{temp2}
\begin{array}{l l}
T_{T,a1} = T_{D,a1} + T_{V,a1} + T_{S,a1} \\
T_{T,a2} = T_{D,a2} + T_{V,a2} + T_{S,a2} \\
\end{array}
\Bigg\} \; \mathsf{with} \; T_{D,a1} \equiv T_{D,a2}
\end{equation*}

and their difference becomes:

\begin{equation}\label{deltaTa1}
\Delta T_T = T_{V,a1} - T_{V,a2} + T_{S,a1} - T_{S,a2}
\end{equation}

Our assumption is that the spillover terms only vary slowly.  $T_V$ will change rapidly on the other hand.  We can remove slow drifts and offsets by calculating differences in a running mean over the time period $t$ around each measurement with $2N$ points:

\begin{equation}\label{Tdc}
T_{DC}(t) = \frac{\sum\limits_{t - N \Delta t}^{t + N \Delta t} T_D + T_V + T_S }{2 N}
\end{equation}

where $\Delta t$ is the time step between each sample, so that the variation between antennae $a1$ and $a2$ becomes defined solely by $T_V$:

\begin{equation}\label{deltaTa}
\Delta T_T \simeq T_{V,a1} - T_{V,a2}
\end{equation}

where these terms are averaged between $t-N\Delta t$ and $t+N\Delta t$. $N$ is the number of separate measurements included in the running mean.  By examining sample data we have determined that the best phase correction is obtained when $N$ is set to be the total number of samples taken for each particular source observation.
As can be seen from Figure \ref{fig:allen_u1} showing the Allan variance, the regime where the WVR noise is gaussian distributed exists from about 1 - 10 minutes with 10 minutes being the optimal observing length to minimise noise. This is well matched to the typical on-source times of 2 - 10 minutes when observing at millimetre wavelengths.

\subsection{Weighting Coefficients}
\label{coefficients}
From these temperature differences ($\Delta T_T$) the water vapour excess between two lines of sight can be determined.  The conversion from temperature differences in each filter and antenna pair to electrical path length $\mathscr{L}_V$ is achieved by deriving a weighting coefficient $C_W$ for each filter.  By modelling the temperatures in the filter frequencies and bandwidths for an assumed standard atmosphere, we determined the ratio of the filter temperature and total wet path for each filter to obtain a water vapour calibration factor $K_f$ for each filter $f$:

\begin{equation}\label{K}
K_f = \frac{T_f}{\mathscr{L}_V}
\end{equation}

Here $\mathscr{L}_V$ is the total wet path for the model atmosphere as given by equation \ref{Lvx}. $T_f$ is the temperature measured over the filter bandwidth for the wet path length, obtained by integrating the Van Vleck -- Weisskopf profile (\cite{Tahmoush:2000uq}) over the respective filter frequency and bandwidth. 

The sensitivity of the wet path length for the range in atmospheric conditions experienced at the ATCA site was examined, with the results shown in Table \ref{tabLv}.  Here we calculated a range of model atmospheres using the ATM code when holding two of the three principal meteorological parameters that determine the wet path length constant (i.e.\ two of $P$, $T$ and $PWV$) and at typical values for the site, while varying the third parameter by the extreme ranges encountered there.  Inspection of the results in Table \ref{tabLv}  shows that it is clearly the $PWV$ variations that dominate changes in the total wet path, with pressure variations having a small effect (about 10\% that of $PWV$) and temperature variations a negligible one.  The effect this has on the calibration factors $K_f$ for each filter is shown in Table \ref{tabKf}.   Our conclusion is that we may use calibration factors calculated for typical values of $T$, $P$ and $PWV$ with small error in the determination of the wet path length, and with the path length variations being dominated by $PWV$ variations.

From the calibration factors and the relative positions on the water line it is then possible to determine the relative weighting coefficients $C_W$ for each filter $f$:

\begin{equation}\label{Cw}
C_{W,f} = \frac{K_{f}^2}{K_{16.5}^2 + K_{18.9}^2 + K_{22.9}^2 + K_{25.5}^2}
\end{equation}

with the normalisation

\begin{equation}\label{SumCw}
\sum_{f=1}^4 C_{W,f} = 1
\end{equation}

The coefficients $C_W$ have been determined for the same range of atmospheric conditions as listed in Tables \ref{tabLv} and \ref{tabKf}.  Their values are shown in Table \ref{tabCw}.  The highest weight coefficient, contributing $\sim 60$\% of the total weight, is for the 22.9\,GHz filter; i.e.\ near the H$_2$O water vapour line centre.  It  varies by less than 4\% from its mean value for the variations encountered in $T$, $P$ and $PWV$\@. The coefficient with the least weight is at 16.5 GHz, the filter least sensitive to the water vapour line, with a weighting of less than 2\%.

For the extraction algorithm, the coefficients for $C_W$ from Table \ref{tabCw}, as determined for an atmosphere with $P$=1013\,hPa, $PWV$=20\,mm and $T$=292\,K are used (i.e. 0.02, 0.09, 0.60, 0.29 respectively, for the four filters).  With the calibration factors $K_f$ (i.e. 0.04, 0.09, 0.23, 0.16, all in $K/mm$), they are applied to the temperature differences obtained from equation \ref{deltaTa}.  This yields the excess path $L_{x,f}$ as determined from each filter:

\begin{equation}\label{grad}
L_{x,f} = \frac{\Delta T_f}{K_f} \; \mathsf{[mm]}
\end{equation}

The weighted sum to obtain $\mathscr{L}_{VX}$ is then built with $f$ = filters 1 through 4 by using the weighting coefficients derived from the model atmosphere in equation \ref{Cw}: 

\begin{equation}\label{LSvx}
\mathscr{L}_{VX} = \sum_{f=1}^{f=4} C_{W,f} \cdot L_{x,f} \; \mathsf{[mm]}
\end{equation}

From this it is trivial to obtain the phase difference $\Delta \Phi$ incurred by the path excess $\mathscr{L}_{VX}$ at observing wavelength $\lambda$ using equation \ref{dPhi}.

\section{Demonstration of Phase Correction}
\label{sec:demonstration}
We demonstrate that the WVRs can track phase variations by comparing the phase measured directly when observing a bright phase calibrator to the phase determined from the WVR signals over the same time period. A phase plot from the unresolved astronomical calibrator source 0537--441 observed at 48.3 GHz is compared to that derived from water vapour radiometers in Figure \ref{fig:phase_plot_61} for the longest baseline used, of 4,500\,m, formed between antennae 1 and 6. We implicitly assume here that the self-calibration undertaken when measuring the phase calibrator tracks the phase perfectly, so that the performance of the WVRs can be assessed by comparing the residuals between it and the calibrator phases to those between the calibrator and the ``interpolated'' phases.  These latter phases represent the best estimates that can be obtained when applying a calibrator phase measurement at the beginning and end of an observation sequence, and then linearly interpolating between them. In Fig.~\ref{fig:phase_plot_61} the calibrator phase fluctuations are shown in black and the WVR derived fluctuations in red.  It is evident that they broadly track each other on long time scales. On short timescales however, there is considerable scatter about the true phase values.  

In Figure \ref{fig:phase_plot_21}  the same time span for the short  baseline of 92\,m between antennae 1 and 2 is shown.  In this case, however, a negligible improvement over the interpolated phase has been obtained.  However, the overall phase noise is also much lower than for the long baselines, as would be expected.

To assess the merit of the WVR corrections we compare the standard deviations of the residual phases between the WVR and calibrator phase and to the residuals between interpolated and calibrator phases.  For the 4,500\,m baseline the residuals are considerably less, with $\sigma = 18.0^{\circ}$ using the WVRs, compared to  47.4$^{\circ}$ for the interpolated residual phases.  On the 92\,m baseline, however, the residuals have barely improved; 9.4$^{\circ}$ for the WVRs compared to 11.0$^{\circ}$ for the interpolated residual phase. 

A second performance indicator we examine compares the interpolated residual phases with the WVR residual phases as a function of baseline length, as plotted in Figure ~\ref{RMS_improvement}.  This shows on long baselines the residual phase error being improved by $\sim 30^{\circ}$, however on short baselines it is worse by $\sim 5^{\circ}$. For practical purposes this latter degradation is negligible because the reduction in correlation efficiency $\epsilon$ (see equation \ref{epsilon}) that corresponds to a 5$^{\circ}$ higher phase noise is less than 6\%.  This can be seen in Figure \ref{fig:epsilon}. On long baselines, in contrast, the correlation efficiency is improved by more than 30\%. The data do suggest, however, that a noise floor limitation of about 10$^{\circ}$, equivalent to a path length of 0.2 mm, in the phase fluctuation at 48 GHz exists which therefore cannot be improved upon using the radiometers.  We can compare this to our theoretical determination of the noise performance by substituting the measurement sensitivity we determined earlier (14 mK) into equations \ref{grad} and \ref{LSvx}. This yields a path difference of 0.08 mm, equivalent to a phase difference of 5$^{\circ}$ at 7 mm. Thus the achieved performance is about a factor of 2 higher than the theoretical limit.

Table \ref{tab:PhaseRMS} lists the resulting standard deviations of the residuals and the correlation efficiencies for all baselines.  As discussed above, the correlation efficiency of the WVR residual phase ($\epsilon_{\mathsf{WVR}}$) is slightly lower than the correlation efficiency of the interpolated residual phase ($\epsilon_{\mathsf{Int}}$) for the short baselines, being reduced by on average 0.02 (but remaining in excess of 0.9).  On the long baselines, however, there is substantial improvement: the correlation efficiency is improved to above 0.9 when the WVR corrections are applied.  For the interpolated residual phase, the correlation efficiency is only about 0.57 for the long baselines.

\section{Summary}
A set of water vapour radiometers (WVRs) has been developed for the Australia Telescope Compact Array that enable the atmospheric phase to be tracked at millimetre-wavelengths while the telescope is observing a source.  These measure the fluctuations of the temperature differences in the telescope beam between each pair of antennae, through each of four 1 GHz wide filters spread across the 22 GHz H$_2$O water vapour line, during the 10\,s cycle time of the telescope.  A set of calibration coefficients have been determined between the path length difference and the temperature difference for each filter based on the properties of a model atmosphere characteristic of the site conditions at Narrabri.  A weighted sum of the four temperature differences then yields the path length difference between each antenna pair, and hence the phase difference at the observing frequency.

The system has been verified by measuring the phase variations between antenna pairs recorded on a bright phase calibrator and compared to the phases determined using the WVRs.  When the phase fluctuations exceed a noise floor (of about $10^{\circ}$ at 48\,GHz, or a path length of $\sim 0.2$\,mm) then the radiometers improve the phase tracking of the telescope.  In practice, depending on the weather conditions, this might yield improvements on long baselines (where phase variations are largest), and slight decorrelations on short baselines.  It will be up to the user to determine when to apply the WVR phase corrections to their astronomical data.  

Further comparison of the WVR-determined phases with phase calibrators under a variety of observing conditions is needed to quantify under which conditions, as a function of baseline and observing frequency, the WVRs may be expected to improve the tracking of phase during the observation of a source.  Such a monitoring program may also be able to yield a set of empirically derived calibration coefficients for each filter, for a variety of weather conditions, that relate the path length difference to the measured temperature fluctuation.  It would be of interest to compare these to the atmospheric model-derived coefficients we have presented in this paper.

\section*{Acknowledgments}
This research was supported under the Australian Research Council’s LIEF grant funding scheme (project number LE0882778). The following institutions have provided funding and support for this project: University of New South Wales, University of Sydney, James Cook University, Swinburne University of Technology, Australia Telescope National Facility (CSIRO/ATNF). The Australia Telescope Compact Array is part of the Australia Telescope National Facility which is funded by the Commonwealth of Australia for operation as a National Facility managed by CSIRO.

We also would like to thank Paul Jones, Brett Hiscock, Philip Edwards, Peter Mirtschin, Christoph Brem and Jock McFee for their contributions integrating, testing and calibrating the WVR systems. And we especially would like to thank the anonymous referee whose comments have helped greatly to improve the clarity of this paper.

\onecolumn[]
\include{tablesfigs}

\bibliographystyle{plainnat}
\bibliography{WVR_REFS}{}
\end{document}

%% file: tablesfigs.tex
\clearpage
\begin{table*}[h]
\centering
\begin{tabularx}{0.8\columnwidth}{llcccc}
\toprule
\multirow{2}{*}{Antenna} & \multirow{2}{*}{Unit} & \multirow{2}{*}{Filter [GHz]} & \multirow{2}{*}{$T_{rec}$ [K]} & Noise Figure [dB] & Noise Floor [dBm]\\
        &      &              &               &      $N_{fig}$    & $N_f$\\
\cmidrule(r){1-6}
& &     16.5 & 305 &  3.12 & --81.58 \\
& &     18.9 & 252 &  2.71 & --81.70 \\
4 & 7 & 22.9 & 321 &  3.23 & --80.72 \\
& &     25.5 & 383 &  3.65 & --80.09 \\
& &       TP & 398 &  3.65 & --70.35 \\
\bottomrule
\end{tabularx}
\caption{The receiver temperatures $T_{rec}$, the noise figure and noise floor for each filter in Unit 7.  TP is the total power channel, which is 10 GHz wide, from 16 to 26 GHz.  For a list of all units, please refer to \cite{Indermuehle:2011nf}}.
\label{tab:noisefigures}
\end{table*}

\clearpage

\begin{table*}[h]
\centering
\begin{tabularx}{0.7\columnwidth}{lccccc}
\toprule
 & 16.5 & 18.9 & 22.9 & 25.5 & Total \\
 & GHz & GHz & GHz & GHz & Power \\
\midrule
Y Factor Aug 2010 &  1.582 &  1.682 &  1.571 &  1.506 & 1.490 \\
Y Factor Nov 2010 &  1.579 & 1.677 & 1.570 & 1.500 & 1.476 \\
Y Factor difference & +0.2\% & +0.3\% & +0.06\% & +0.4\% & +0.9\% \\
\midrule
$T_{Rec}$ K Aug 2010 &  304.1 & 247.9 & 311.4 & 360.9 & 375.4 \\
$T_{Rec}$ K Nov 2010 &  306.0 & 250.3 & 311.9 & 366.6 & 388.3 \\
$T_{Rec}$ difference & 1.0\% & 1.0\% & 1.0\% & --1.5\% & --3.4\% \\
\bottomrule
\end{tabularx}
\caption{Gain comparisons for calibrations executed on Unit 7 in August and November 2010. }
\label{tab:gaincals}
\end{table*}

\clearpage

\begin{table*}[h]
\centering
\begin{tabularx}{0.63\columnwidth}{lcccccc}
\toprule
Antenna & Unit & Filter [GHz] & $\tau$ & $\tau_{\mathsf{err}}$ & $T_{S}$[K] & $T_{S,\mathsf{err}}$[K] \\
\midrule
\multirow{4}{*}{CA01} & \multirow{4}{*}{1} & 16.5  &   0.04 &   0.001 &  --2.4 &   0.50\\
 & & 18.9  &   0.07 &   0.001 &  --3.7 &   0.57\\
 & & 22.9  &   0.20 &   0.003 &  --6.5 &   0.95\\
 & & 25.5  &   0.13 &   0.002 &  --5.6 &   0.83\\
\midrule
\multirow{4}{*}{CA02} & \multirow{4}{*}{2} & 16.5 &   0.04 &   0.001 &  --6.9 &   0.48 \\
 & & 18.9 &   0.07 &   0.002 &  --6.6 &   0.56 \\
 & & 22.9 &   0.21 &   0.005 & --12.1 &  1.60 \\
 & & 25.5 &   0.13 &   0.004 & --11.3 &   1.14\\
\midrule
\multirow{4}{*}{CA03} & \multirow{4}{*}{3} & 16.5 &   0.05 &   0.001 &  --7.0 &   0.47 \\
 & & 18.9 &   0.08 &   0.002 &  --9.0 &   0.84 \\
 & & 22.9 &   0.23 &   0.008 & --14.1 &   2.26 \\
 & & 25.5 &   0.13 &   0.004 &  --8.8 &   1.37 \\
\midrule
\multirow{4}{*}{CA04} & \multirow{4}{*}{7} & 16.5 &   0.05 &   0.001 & --10.3 &   0.43 \\
 & & 18.9 &   0.08 &   0.002 & --11.2 &   0.72 \\
 & & 22.9 &   0.25 &   0.009 & --21.3 &   2.35 \\
 & & 25.5 &   0.15 &   0.003 & --15.1 &   1.12 \\
\midrule 
\multirow{4}{*}{CA05} & \multirow{4}{*}{5} & 16.5 &   0.04 &   0.001 &  --7.1 &   0.39 \\
 & & 18.9 &   0.08 &   0.002 &  --8.5 &   0.69 \\
 & & 22.9 &   0.23 &   0.007 & --13.7 &   1.97 \\
 & & 25.5 &   0.13 &   0.004 &  --9.4 &   1.36 \\
\midrule
\multirow{4}{*}{CA06} & \multirow{4}{*}{6} & 16.5 &   0.04 &   0.001 &  --4.6 &   0.23 \\
 & & 18.9 &   0.07 &   0.001 &  --3.7 &   0.45 \\
 & & 22.9 &   0.19 &   0.004 &  --3.8 &   1.37 \\
 & & 25.5 &   0.11 &   0.001 &  --2.7 &   0.75 \\
\bottomrule
\end{tabularx}
\caption{The zenith opacity $\tau$ and spillover temperature $T_S$, with their corresponding errors $\tau_{\mathsf{err}}$ and $T_{S,\mathsf{err}}$ for each antenna, unit and filter, as determined from the fits to the data shown in Fig.~\ref{fig:skydip} and eqn.~\ref{spillover}.}
\label{tab:spillover}
\end{table*}

\clearpage

\begin{table*}[h]
\centering
\begin{tabularx}{0.5\columnwidth}{lcr}
\toprule
Atmosphere & Variable & Total Wet Path $\mathscr{L}_v$ \\
\midrule
\multirow{3}{*}{\begin{minipage}{0.5cm}P=1013~hPa T=292~K \end{minipage}} & PWV 30 mm	& 181.2 mm\\
& PWV 1 mm	& 6.0 mm\\
& Spread & 87.6 mm\\
\midrule
\multirow{3}{*}{\begin{minipage}{0.5cm}T=292~K PWV=20~mm\end{minipage}} & 1030 hPa	& 120.8 mm\\
& 960 hPa	& 120.8 mm\\
& Spread & 0.0 mm\\
\midrule
\multirow{3}{*}{\begin{minipage}{0.5cm}P=1013~hPa PWV=20~mm\end{minipage}} & 270 K	& 130.6 mm\\
& 310 K	& 113.9 mm\\
& Spread & 8.3 mm\\
\bottomrule
\end{tabularx}
\caption{Sensitivity of the total wet path, $\mathscr{L}_v$, to variations in atmospheric conditions. In the first entry, $PWV$ is varied between extreme values encountered, while keeping the pressure $P$ and temperature $T$ constant and at typical values for the site. The sensitivity (``spread'') is shown in the third row and is half the range; in this case being a wet path $\mathscr{L}_v$ variation of 87.6\,mm. In the second entry $P$ is varied, showing no effect on $\mathscr{L}_v$.  In the third entry, $T$ is varied, with a variation in $\mathscr{L}_v$ of less than 10\% that occurring when varying $PWV$.}
\label{tabLv}
\end{table*}

\clearpage

\begin{table*}[h]
\centering
\begin{tabularx}{0.8\columnwidth}{lccccc}
\toprule
 & & \multicolumn{4}{c}{Water Vapour Calibration Factor $K_f$ [K/mm]} \\
 \cmidrule(r){3-6}
Atmosphere & Variable & 16.5 GHz & 18.9 GHz & 22.9 GHz & 25.5 GHz \\
\midrule
\multirow{3}{*}{\begin{minipage}{0.5cm}P=1013~hPa T=292~K \end{minipage}} & PWV 30 mm	& 0.044 & 0.091 & 0.232 & 0.164 \\
& PWV 1 mm	& 0.043 & 0.091 & 0.259 & 0.169 \\
& Spread & 0.001 & 0.000 & 0.010 & 0.003 \\
\midrule
\multirow{3}{*}{\begin{minipage}{0.5cm}T=292~K PWV=20~mm\end{minipage}} & 1030 hPa	& 0.044 & 0.092 & 0.239 & 0.166 \\
& 960 hPa	& 0.042 & 0.089 & 0.248 & 0.162\\
& Spread & 0.001 & 0.001 & 0.005 & 0.001 \\
\midrule
\multirow{3}{*}{\begin{minipage}{0.5cm}P=1013~hPa PWV=20~mm\end{minipage}} & 270 K	& 0.041 & 0.083 & 0.207 & 0.150 \\
& 310 K	& 0.045 & 0.098 & 0.269 & 0.177 \\
& Spread & 0.007 & 0.010 & 0.031 & 0.014 \\
\bottomrule
\end{tabularx}
\caption{The water vapour calibration factor $K_f$ for each filter under a variety of atmospheric conditions.  The spread is half the range in path length for the variation in the relevant variable (column 2).  The largest spread occurs in the 22.9 GHz filter, at the peak of the water vapour line.}
\label{tabKf}
\end{table*}

\clearpage

\begin{table*}[h]
\centering
\begin{tabularx}{0.8\columnwidth}{lccccc}
\toprule
 & & \multicolumn{4}{c}{Weighting Coefficient $C_W$} \\
 \cmidrule(r){3-6}
Atmosphere & Variable & 16.5 GHz & 18.9 GHz & 22.9 GHz & 25.5 GHz \\
\midrule
\multirow{3}{*}{\begin{minipage}{0.5cm}P=1013~hPa T=292~K \end{minipage}} & PWV 30 mm	& 0.0213 & 0.0911 & 0.5929 & 0.2946 \\
& PWV 1 mm	& 0.0172 & 0.0791 & 0.6353 & 0.2685 \\
& Spread & 0.0021	& 0.006 & 0.0212 & 0.0131 \\
\midrule
\multirow{3}{*}{\begin{minipage}{0.5cm}T=292~K PWV=20~mm\end{minipage}} & 1030 hPa	& 0.0205 & 0.0886 & 0.5998 & 0.2912 \\
& 960 hPa	& 0.0179 & 0.0818 & 0.6304 & 0.2698 \\
& Spread & 0.0013 & 0.0034 & 0.0153 & 0.0107 \\
\midrule
\multirow{3}{*}{\begin{minipage}{0.5cm}P=1013~hPa PWV=20~mm\end{minipage}} & 270 K	& 0.0228 & 0.0930 & 0.5798 & 0.3045 \\
& 310 K	& 0.0179 & 0.0825 & 0.6272 & 0.2723 \\
& Spread & 0.0025 & 0.0053 & 0.0237 & 0.0161 \\
\bottomrule
\end{tabularx}
\caption{The weighting coefficients $C_W$ for each filter under a variety of atmospheric conditions.  The spread is half the range in path length for the variation in the relevant variable (column 2).  The largest spread occurs in the 22.9 GHz filter and amounts to an uncertainty of 3.5\% in its value.}
\label{tabCw}
\end{table*}

\clearpage

\begin{table*}[h]
\centering
\begin{tabularx}{0.7\columnwidth}{lccccc}
\toprule
 & \multicolumn{2}{c}{Interpolated} & \multicolumn{2}{c}{WVR Corrected} & Improvement in \\
\cmidrule(r){2-3}
\cmidrule(r){4-5}
Baseline [m] & $\sigma_{\mathsf{Int}}$ & $\epsilon_{Int}$ & $\sigma_{\mathsf{WVR}}$ & $\epsilon_{WVR}$ & correlation efficiency $\Delta\epsilon$ \\
\midrule
92   & 11.0 & 0.96 & 9.4 & 0.97 & 0.01 \\
230  & 10.9 & 0.96 & 10.3 & 0.97 & 0.00 \\
138  & 3.4 & 1.00 & 9.5 & 0.97 & --0.02 \\
144  & 5.1 & 0.99 & 13.5 & 0.95 & --0.05 \\
82   & 7.5 & 0.98 & 11.5 & 0.96 & --0.02  \\
132  & 7.3 & 0.98 & 11.7 & 0.96 & --0.03 \\
247  & 4.1 & 0.94 & 16.6 & 0.92 & --0.02 \\
216  & 5.7 & 0.99 & 15.5 & 0.93 & --0.06 \\
240  & 4.6 & 0.99 & 15.3 & 0.93 & --0.06  \\
138  & 10.6 & 0.97 & 13.3 & 0.95 & --0.02\\
4500 & 47.4 & 0.50 & 18.0 & 0.91 & 0.40 \\
4408 & 41.9 & 0.59 & 17.5 & 0.91 & 0.32 \\
4270 & 41.3 & 0.60 & 16.2 & 0.92 & 0.33 \\
4378 & 45.8 & 0.53 & 15.9 & 0.93 & 0.40 \\
4383 & 39.5 & 0.62 & 12.2 & 0.96 & 0.33\\
\bottomrule
\end{tabularx}
\caption{Comparison of figures of merit for each baseline: Listed are the standard deviations $\sigma$ for the interpolated residual phases and for the WVR residual phases, along with their respective correlation efficiencies $\epsilon$. }
\label{tab:PhaseRMS}
\end{table*}

\clearpage

\begin{figure}[h]
\centering
\includegraphics[width=1.0\columnwidth]{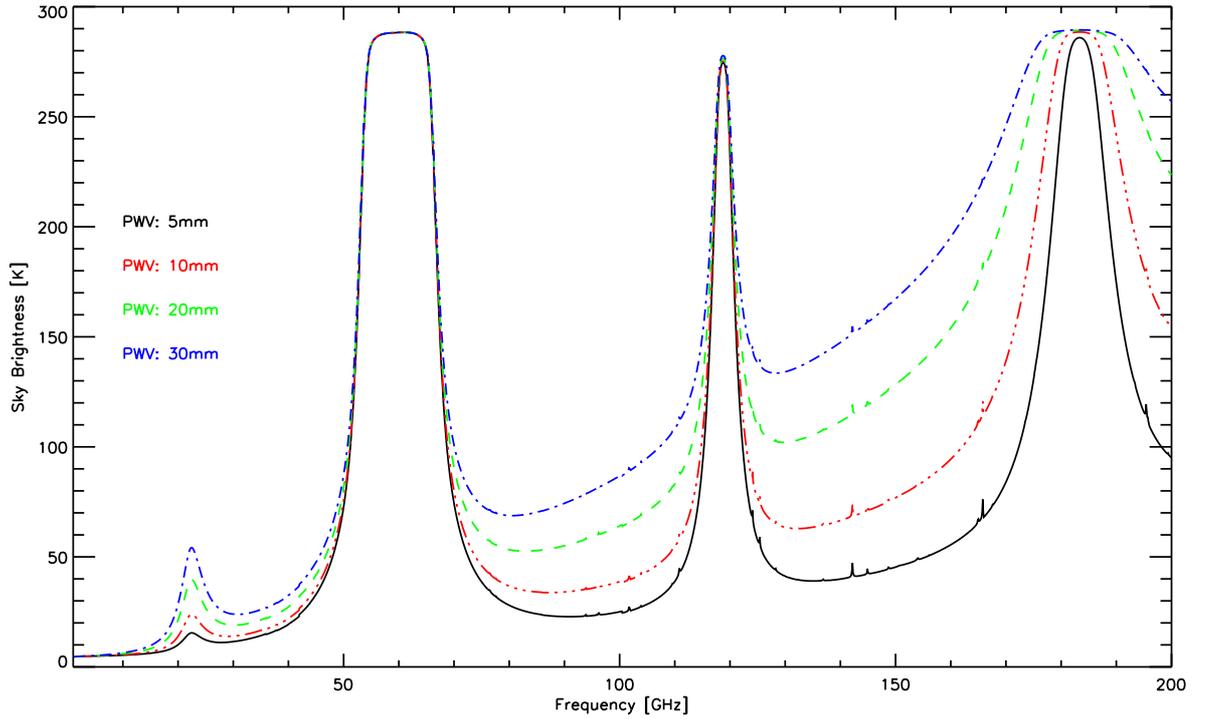}
\caption{The atmospheric emission from 1\,GHz to 200\,GHz for a range of PWV conditions, modelled using the \cite{Pardo:2001fk} ATM model.  Variations in precipitable water vapour (PWV) are shown from 1\,mm to 30\,mm with a constant atmosphere at 235\,m elevation, 1000\,hPa pressure and a temperature of\,290\,K, typical of the Narrabri site. The features at 22.2\,GHz and 183.3\,GHz are the prominent water vapour lines.  The 60\,GHz and 118\,GHz features are caused by O\textsubscript{2}.  It is immediately evident that for the PWV values of 5\,mm to 30\,mm encountered at the ATCA site in Narrabri, only the 22.3\,GHz line is a viable candidate as the 183.3\,GHz line is completely saturated.  For locations where the PWV falls below 2\,mm, the 183.3\,GHz line is the better choice.  Note also that the continuum increases as $\approx \nu^2$, arising from the liquid water contribution.}
\label{figure1}
\end{figure}

\clearpage

\begin{figure}[h]
\centering
\includegraphics[width=1\columnwidth]{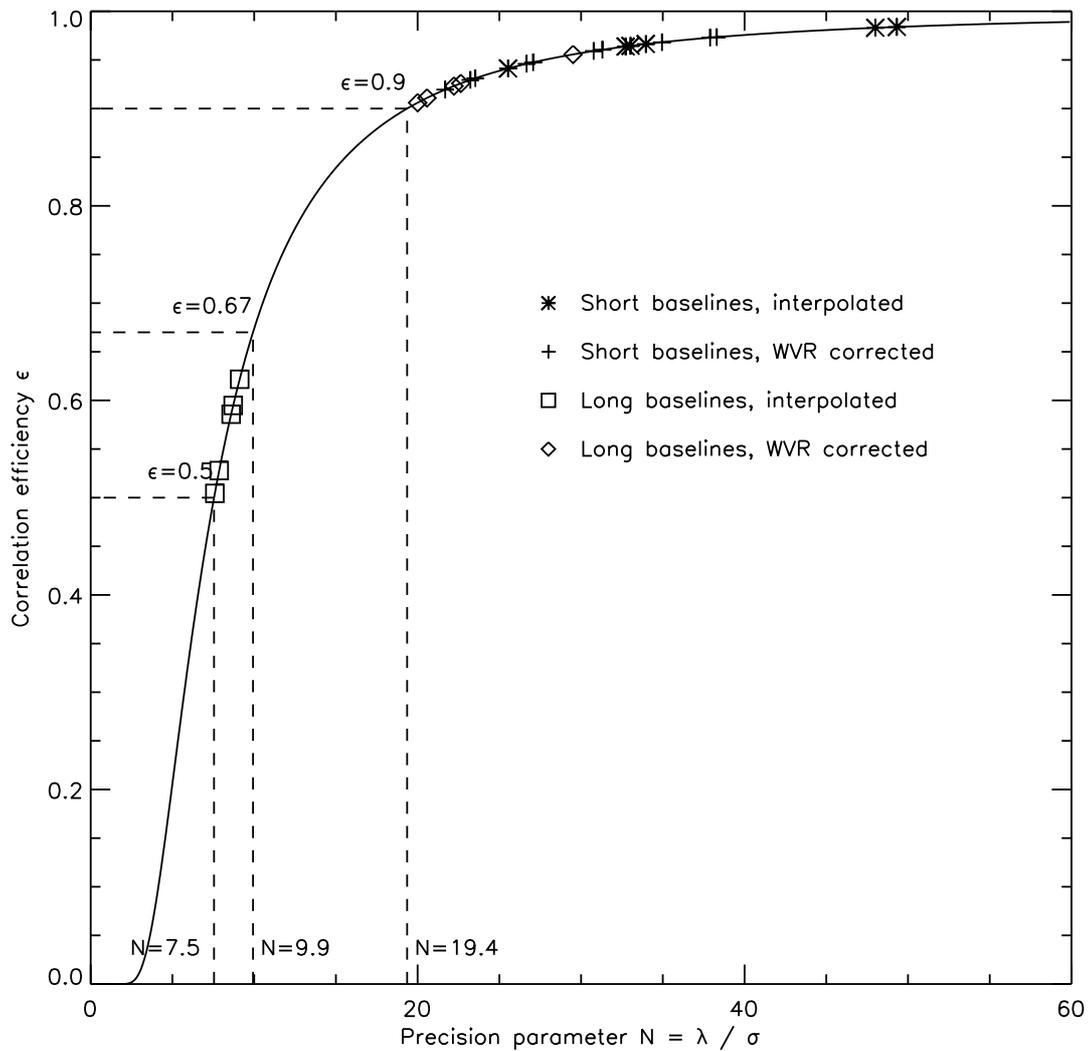}
\caption{Precision parameter, $N=\lambda/\sigma$ vs. the resulting correlation efficiency, $\epsilon$: The solid line plots the Ruze formula (see equation \ref{Ruze}).  Over-plotted are shown the resulting correlation efficiency from the measured phase noise in test observations (see \S\ref{sec:demonstration}).  Stars show short baselines with interpolated data, where correlation efficiency is generally better than 0.95. Plus signs show these baselines with WVR corrections applied.  While this results in added noise, the lowered correlation efficiency remains above 0.9.  On the long baselines, the WVR improvements are substantial: without corrections, efficiencies are about 0.65 (squares), with corrections better than 0.9 (diamonds).}
\label{fig:epsilon}
\end{figure}

\clearpage

\begin{figure}[h]
\includegraphics[width=\textwidth]{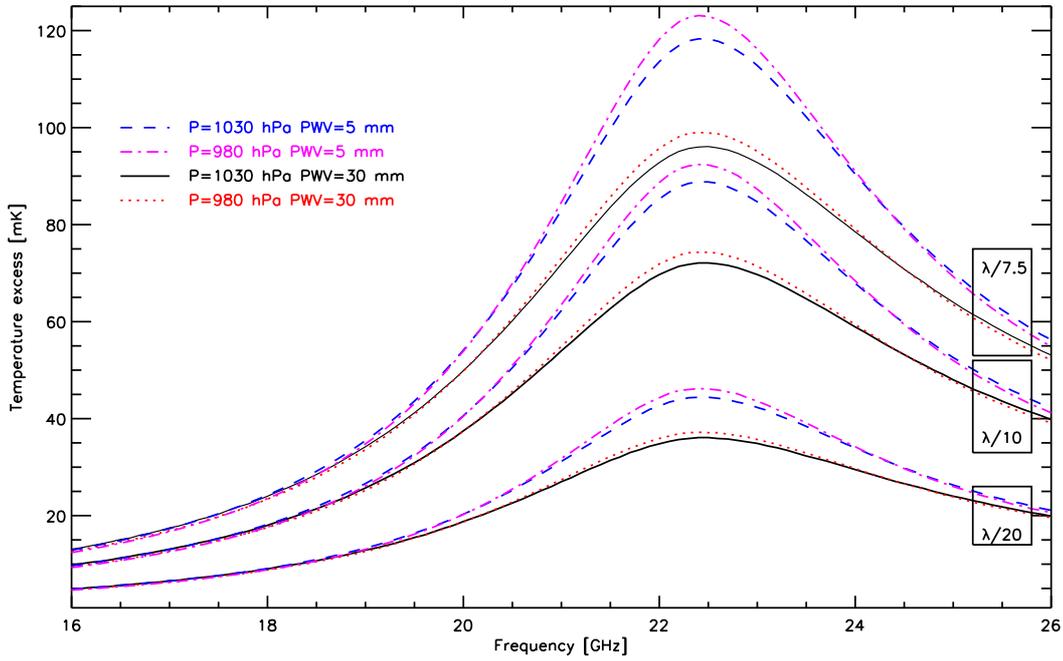}
\caption{The temperature excess for a difference in PWV equivalent to a path difference of 1/7.5, 1/10 and 1/20 of the observing wavelength, 3 mm, for four different atmospheric ATM models as labelled (each for a temperature of 292 K). The boxes labelling the right hand side show which models relate to which path difference. For example, the solid black line corresponds to a PWV of 30 mm and a pressure of 1030 hPa. The intersect with the $y$-axis at 16 GHz shows the required temperature sensitivity for each path difference: to correct to $\lambda/7.5$, 14 mK is required, to correct to $\lambda/10$, 12 mK is required and to correct to $\lambda/20$, 5 mK.}
\label{deltaPWV}
\end{figure}

\clearpage

\begin{figure}[h]
\centering
\includegraphics[width=1\columnwidth]{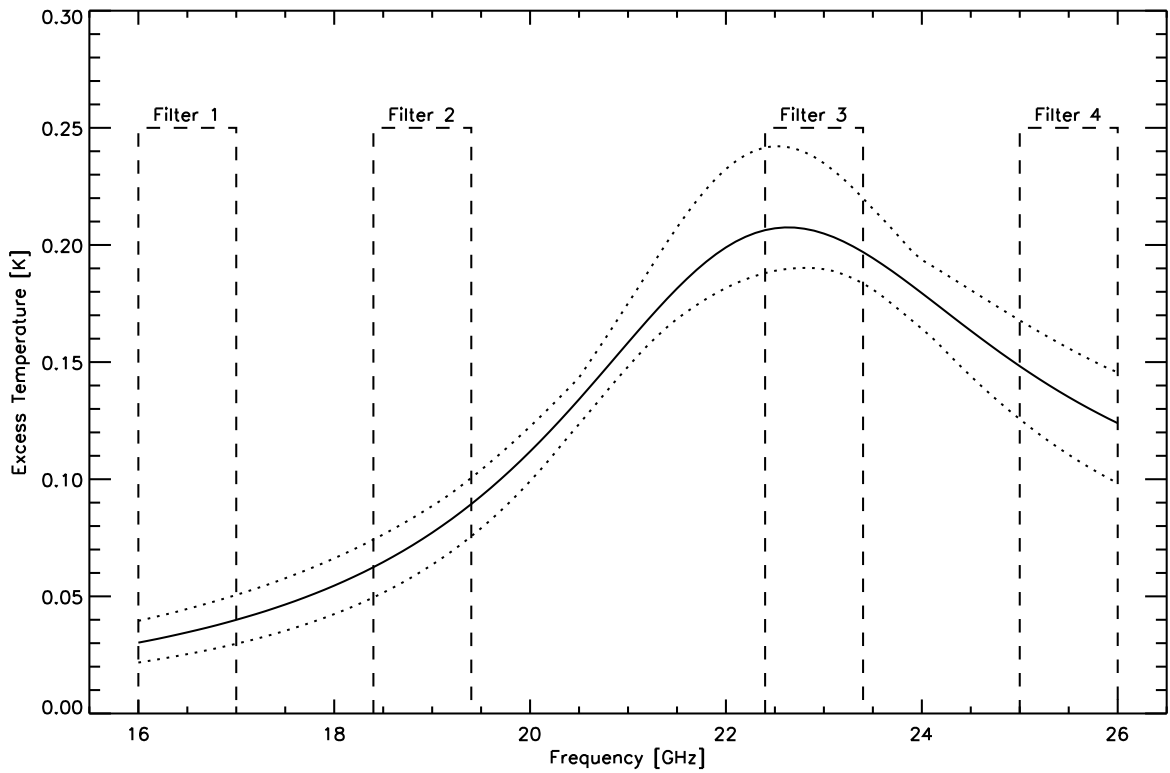}
\caption{Filter placement over the 22.2 GHz water line: Shown are the overall water line excess variations for a 10 year period using radio sonde data as input for the model and when there is a 1\,mm path length difference between the signals measured by two radiometers. The dotted lines represent the maxima/minima values encountered in the entire data set.}
\label{fig:filterplacement}
\end{figure}

\clearpage 

\begin{figure}[h]
\centering
\includegraphics[width=1\columnwidth]{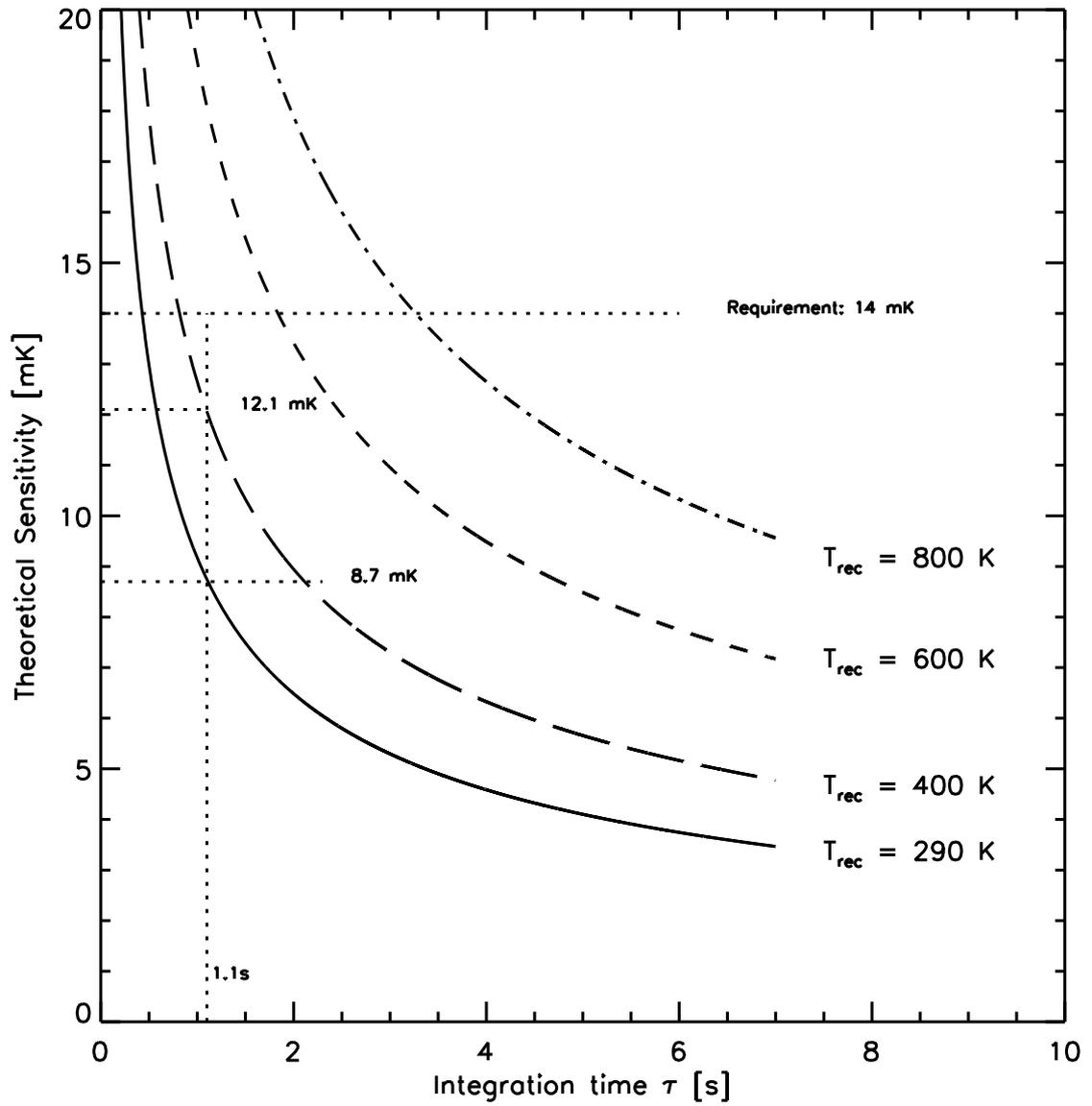}
\caption{The theoretical sensitivity of a radiometer with 1 GHz bandwidth as given by equation \ref{deltaT}, calculated for receiver temperatures of 290 K, 400 K, 600 K, 800 K and 1600 K respectively. At a $T_{rec}$ of 400 K, 1.1s integration time will yield 12.1 mK sensitivity, which is slightly better than the required 14 mK.}
\label{fig:sensitivity}
\end{figure}

\clearpage



\begin{figure}[h]
\centering
\includegraphics[width=1\columnwidth]{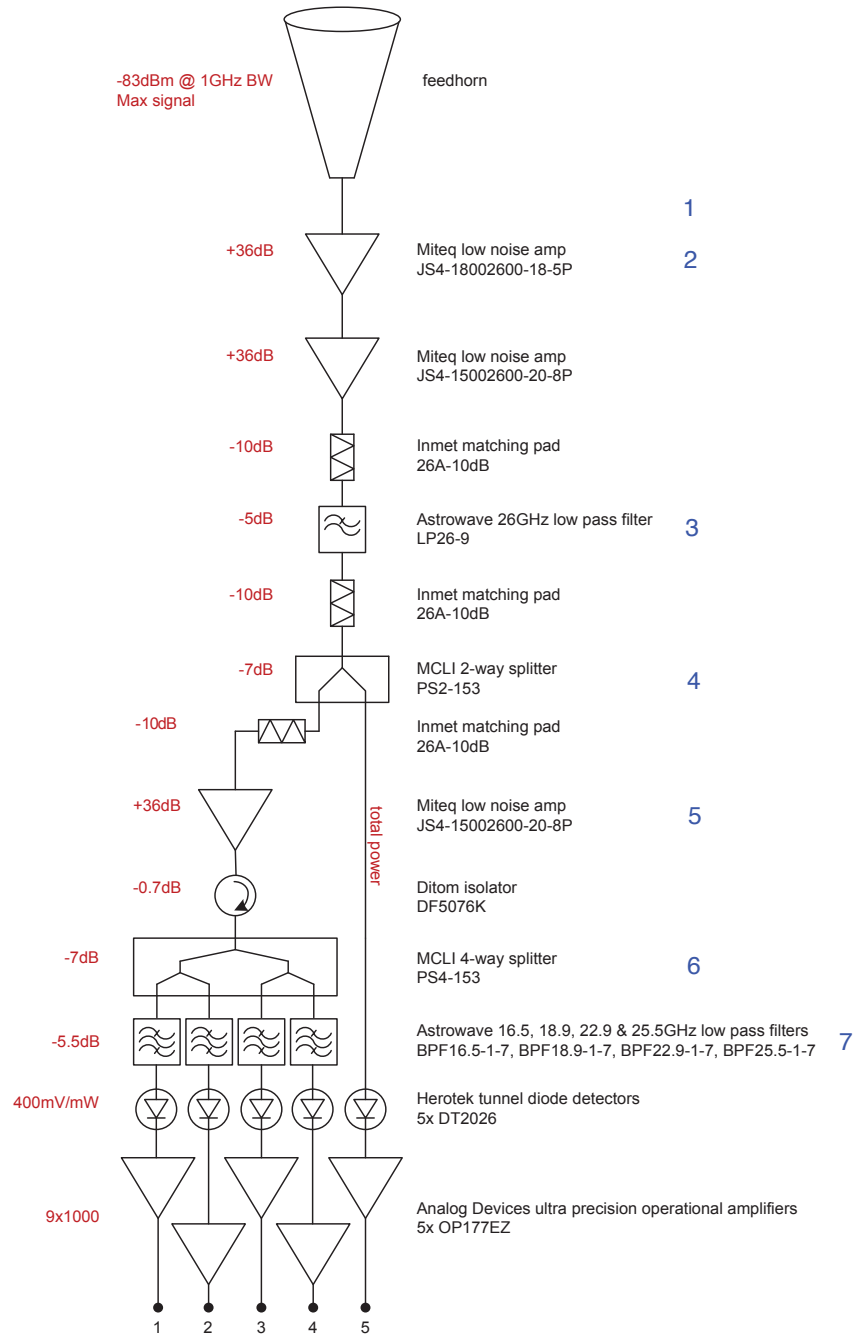}
\caption{A schematic diagram of the WVR kindly provided by Christoph Brem (CASS Narrabri). The dB numbers on the lefthand side show the signal loss (or gain) at each stage along the signal path. The blue figures to the right are used as stage references in the text.  }
\label{fig:schematics}
\end{figure}

\clearpage

\begin{figure}[h]
\centering
\includegraphics[width=1\columnwidth]{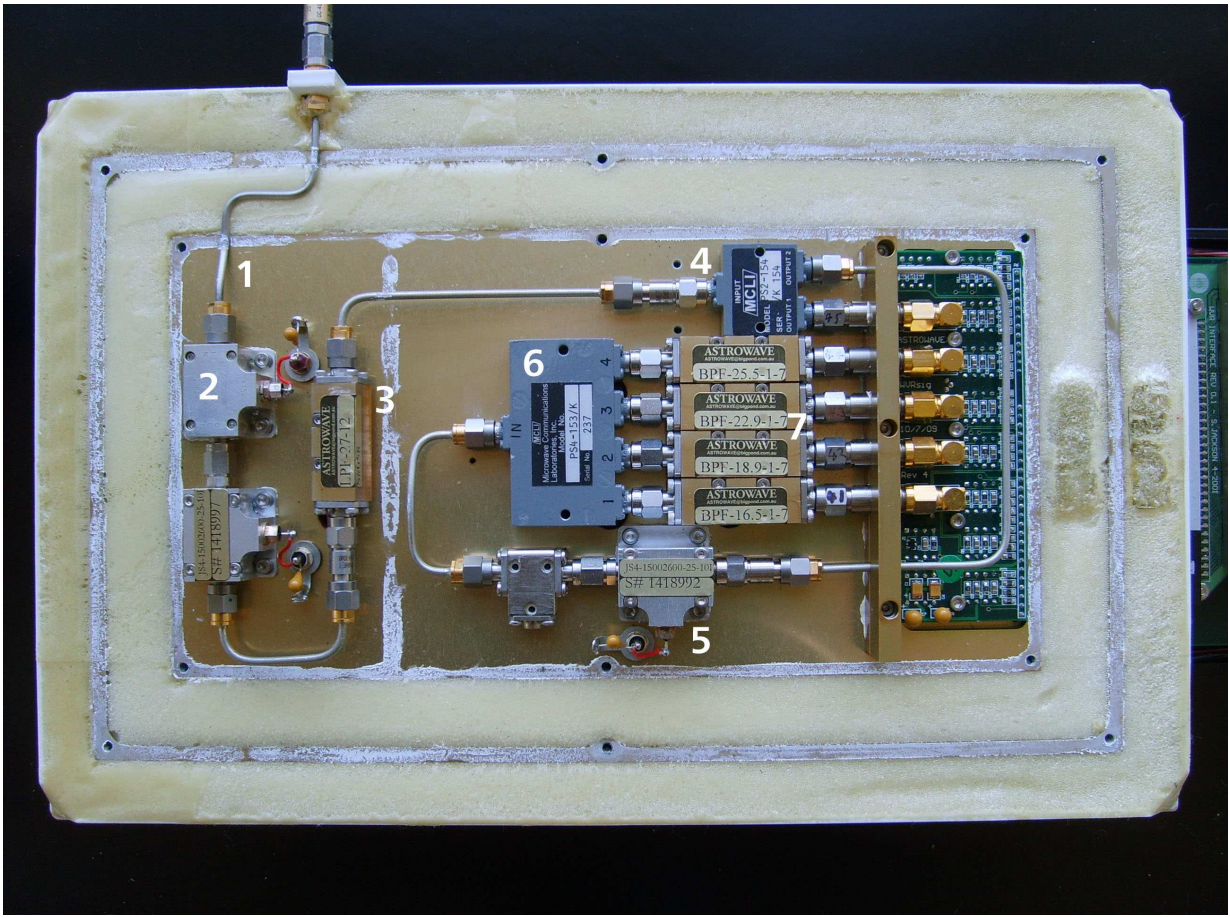}
\caption{A photograph of the RF plate of Unit 4.  The foam insulating material is visible around the unit and the components can be identified by comparing to Figure \ref{fig:schematics}.}
\label{fig:unit4photo}
\end{figure}

\clearpage

\begin{figure}[h]
\centering
\includegraphics[width=1\columnwidth]{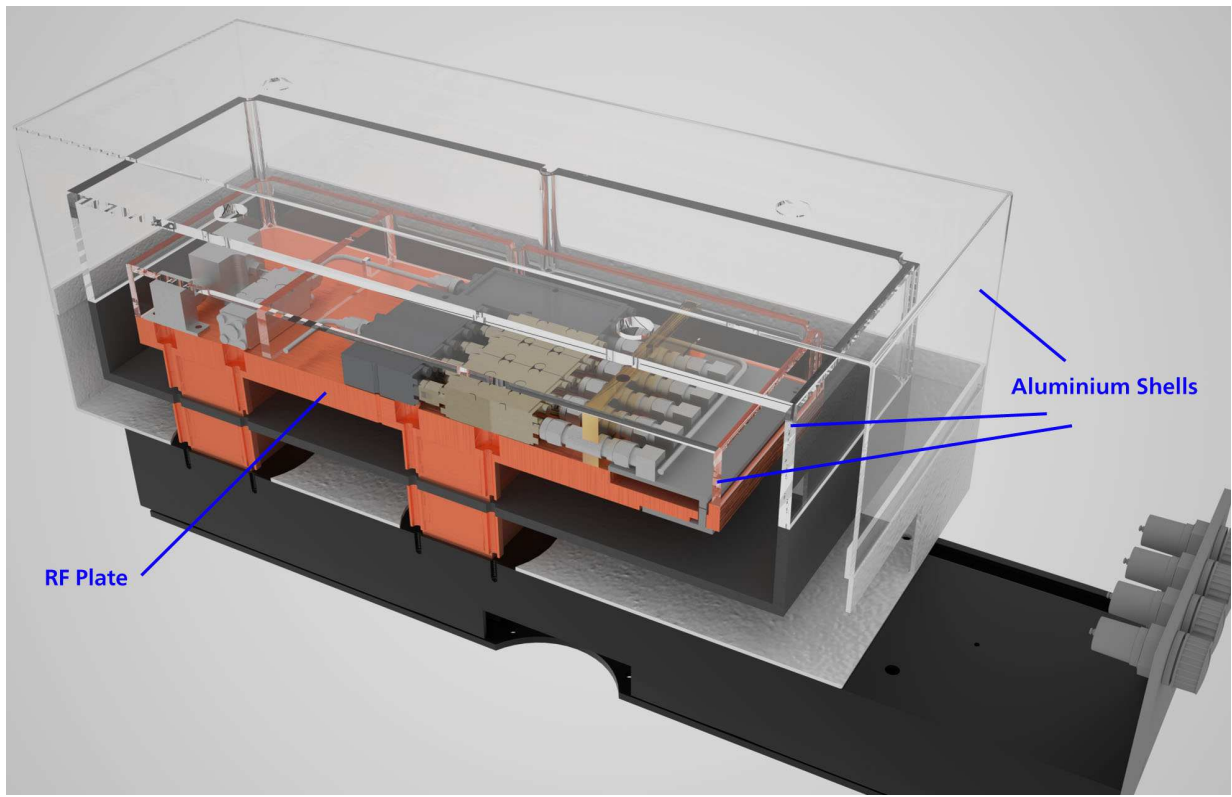}
\caption{Cross section render through one of the WVRs.  In the centre is the RF plate with components visible.  The RF plate itself is shielded inside a first aluminium enclosure, which itself is surrounded by foam (not shown in the render).  Then a second aluminium shell surrounds this with more foam on the outside and finally the outside shell which is not thermally controlled.  The outside is however painted white to minimise thermal energy uptake through solar radiation when pointing the antennas near the Sun.  Visualised by the author using Maxwell Render (\cite{Maxwell:2011fk}) based on an AutoCAD model. }
\label{fig:section1}
\end{figure}

\clearpage

\begin{figure}[h]

\centering
\includegraphics[width=1\columnwidth]{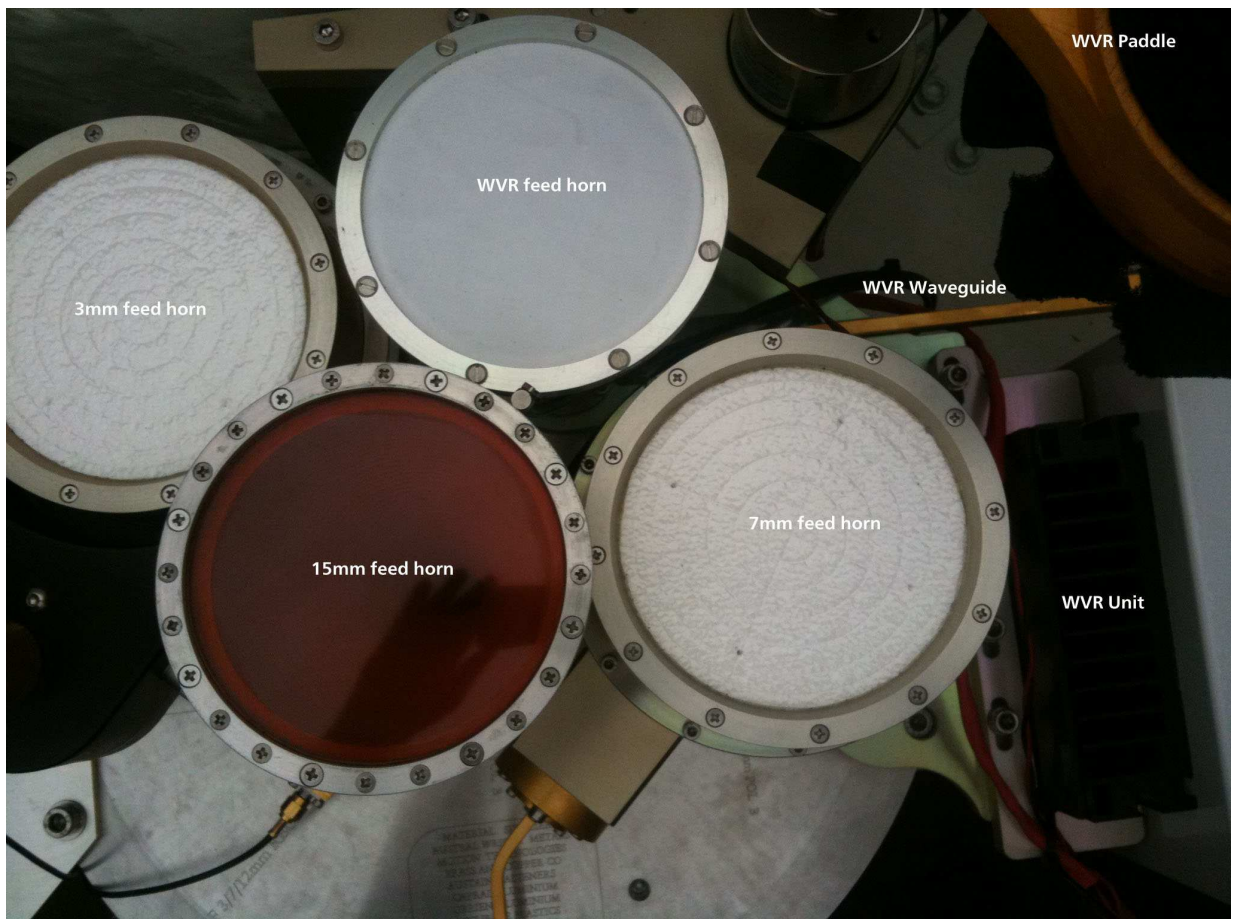}
\caption{Top view of the millimetre dewar with all the feeds visible.  Clockwise from the top: WVR feed with gold coloured waveguide leading the sky signal to the WVR box mounted on the right hand side of the millimetre package.  Next is the 7\,mm feed, then follows 15\,mm and lastly to the left is the 3\,mm feed. }
\label{fig:topdewar}
\end{figure}

\clearpage

\begin{figure}[h]
\centering
\includegraphics[width=1\columnwidth]{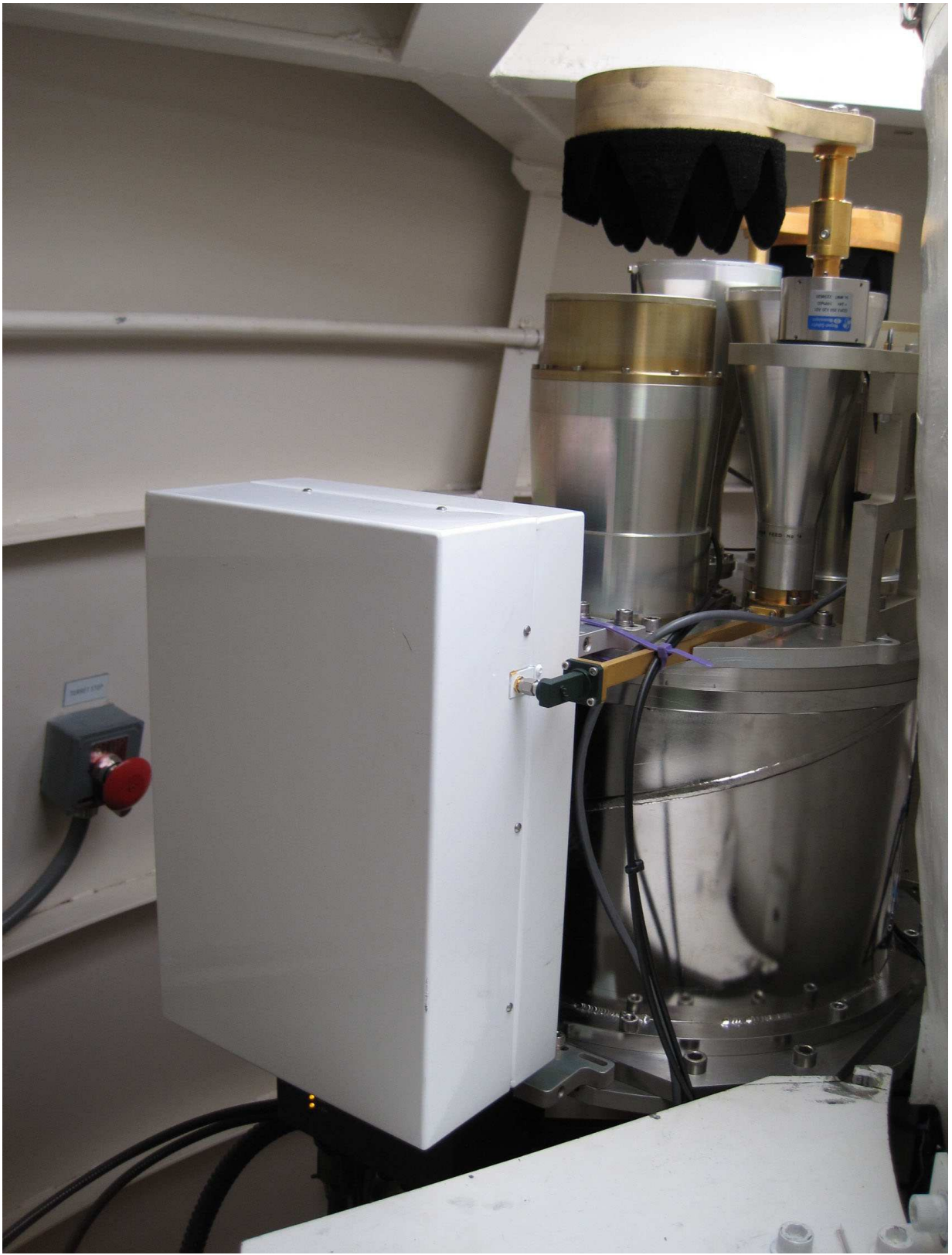}
\caption{A side view of the millimetre package with the WVR mounted on the side (the white box).  The WVR paddle, feed horn as well as waveguide can be clearly seen.  The WVR paddle is in the same position as in Figure \ref{fig:topdewar}, i.e.  not obstructing any of the feed horns.  Photo by Peter Mirtschin in April 2011. }
\label{fig:sideview}
\end{figure}

\clearpage
\begin{figure}[h]
\centering
\includegraphics[width=1\columnwidth]{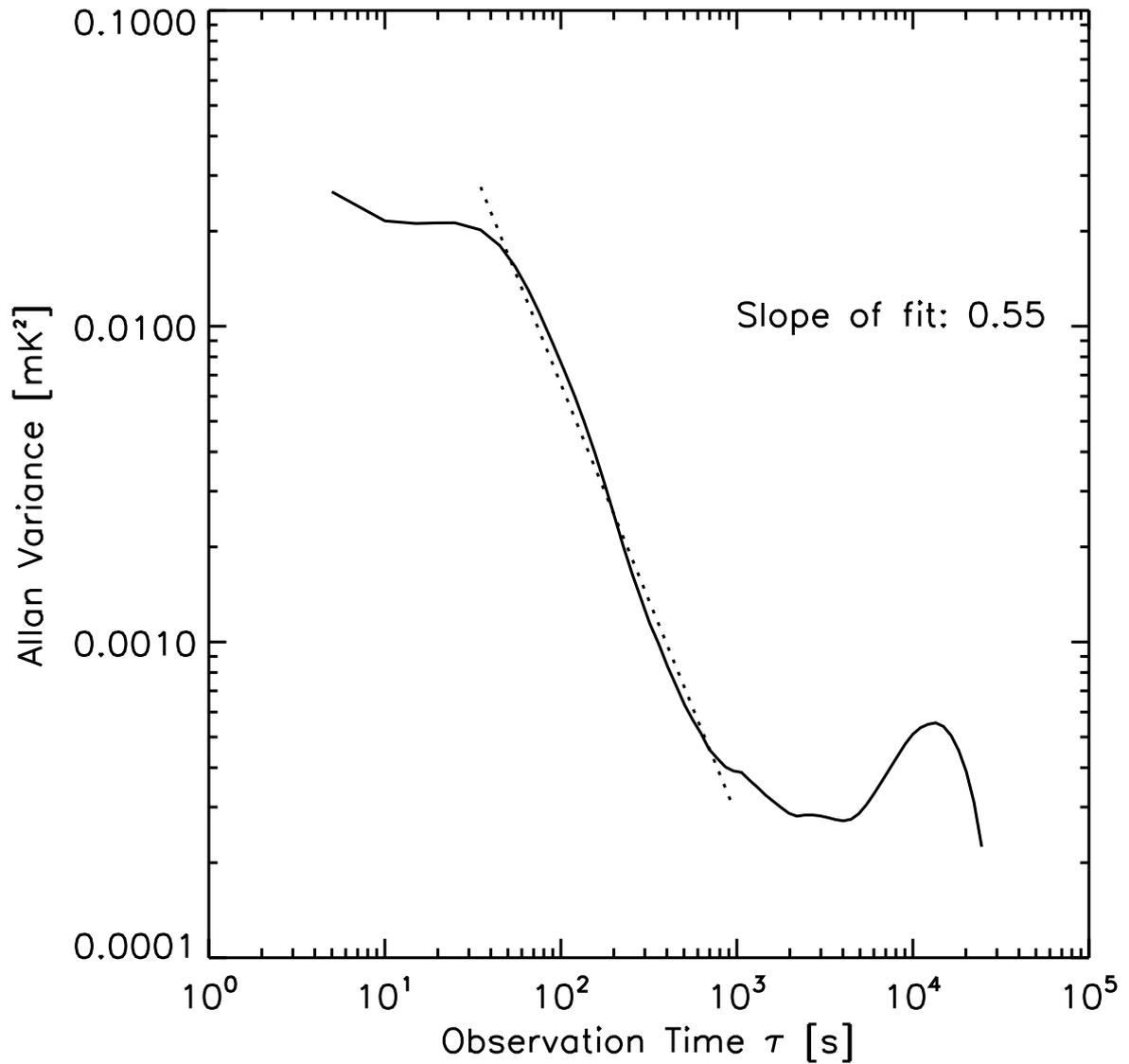}
\caption{The Allan variance of the RF Plate temperature control point for Unit 6 on antenna ca06 in the evening of August 23 2011. The noise of the shortest observation time (the single point noise) is 0.4 mK and therefore already exceeds the required temperature stability of 1 mK. The temperature variations are gaussian noise dominated between 60 to 600 seconds (1 - 10 minutes), indicating there are no systematic causes to that noise other than random fluctuations. On longer timescales, low frequency noise increases the Allan variance. The best integration timescales therefore are less than an hour. }
\label{fig:allen_u1}
\end{figure}

\clearpage

\begin{figure}[h]
\centering
\includegraphics[width=1\columnwidth]{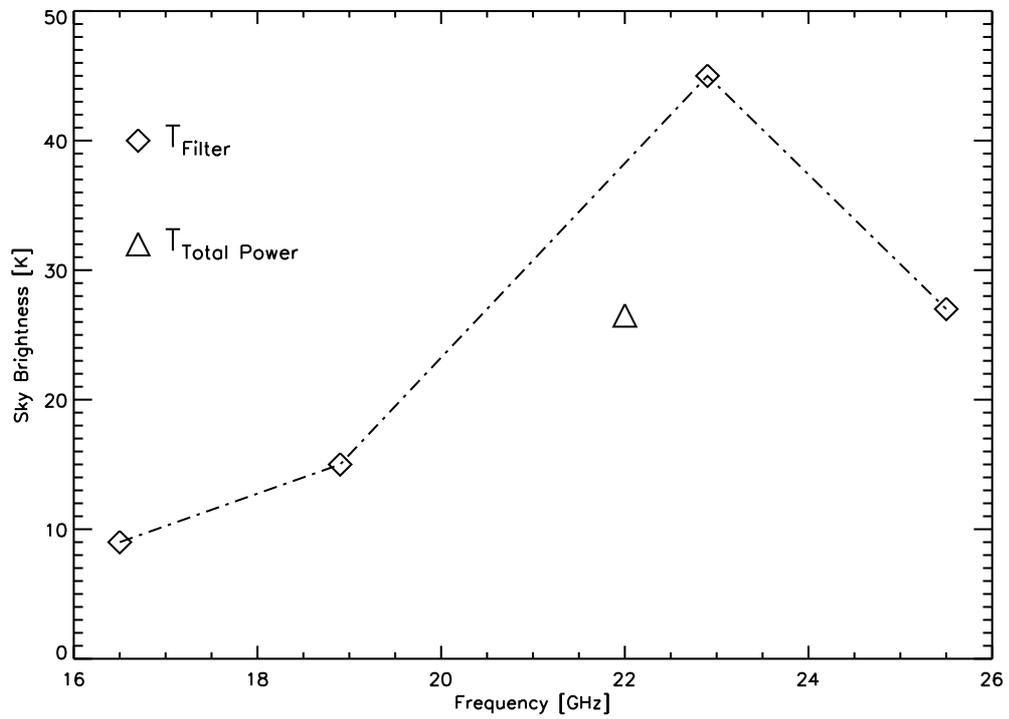}
\caption{Zenith sky spectrum for ca01 (unit 1).  The line peaks in the 22.9\,GHz channel and exceeds the $\nu^2$ continuum level (indicated by the other three channels) by $\sim 20$\,K\@.  The point for Total Power (triangle) shows the sky brightness measured across the entire 16--24\,GHz band pass. These data were taken on March 7 2011 in clear sky conditions.}
\label{unit1_zenith_spectrum}
\end{figure}

\clearpage

\begin{figure}[h]
\centering
\includegraphics[width=1\columnwidth]{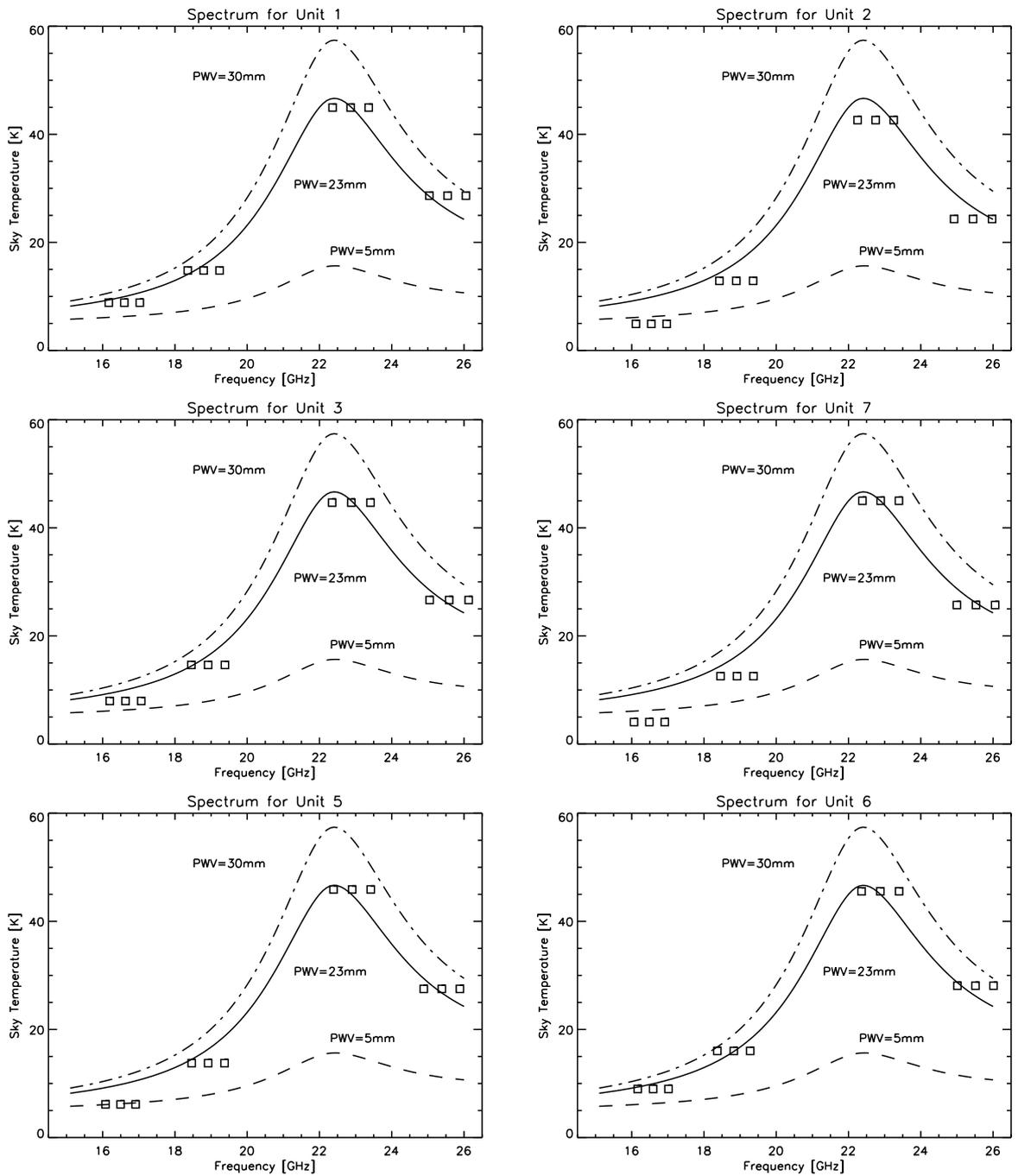}
\caption{Sky brightness measurements for each Unit as determined during calibration in March 2011.  The squares show the sky temperatures in the 4 filters, with the three boxes for each filter indicating their centres and widths. The three lines are for PWV values of 5\,mm (the lowest PWV value recorded at Narrabri), 30\,mm (the amount of PWV where millimetre observing is discountinued) and 23\,mm (as determined from the data at the time of observation). The profiles were calculated using the ATM code.}
\label{fig:spectrum}
\end{figure}

\clearpage

\begin{figure}[h]
\centering
\includegraphics[width=1\columnwidth]{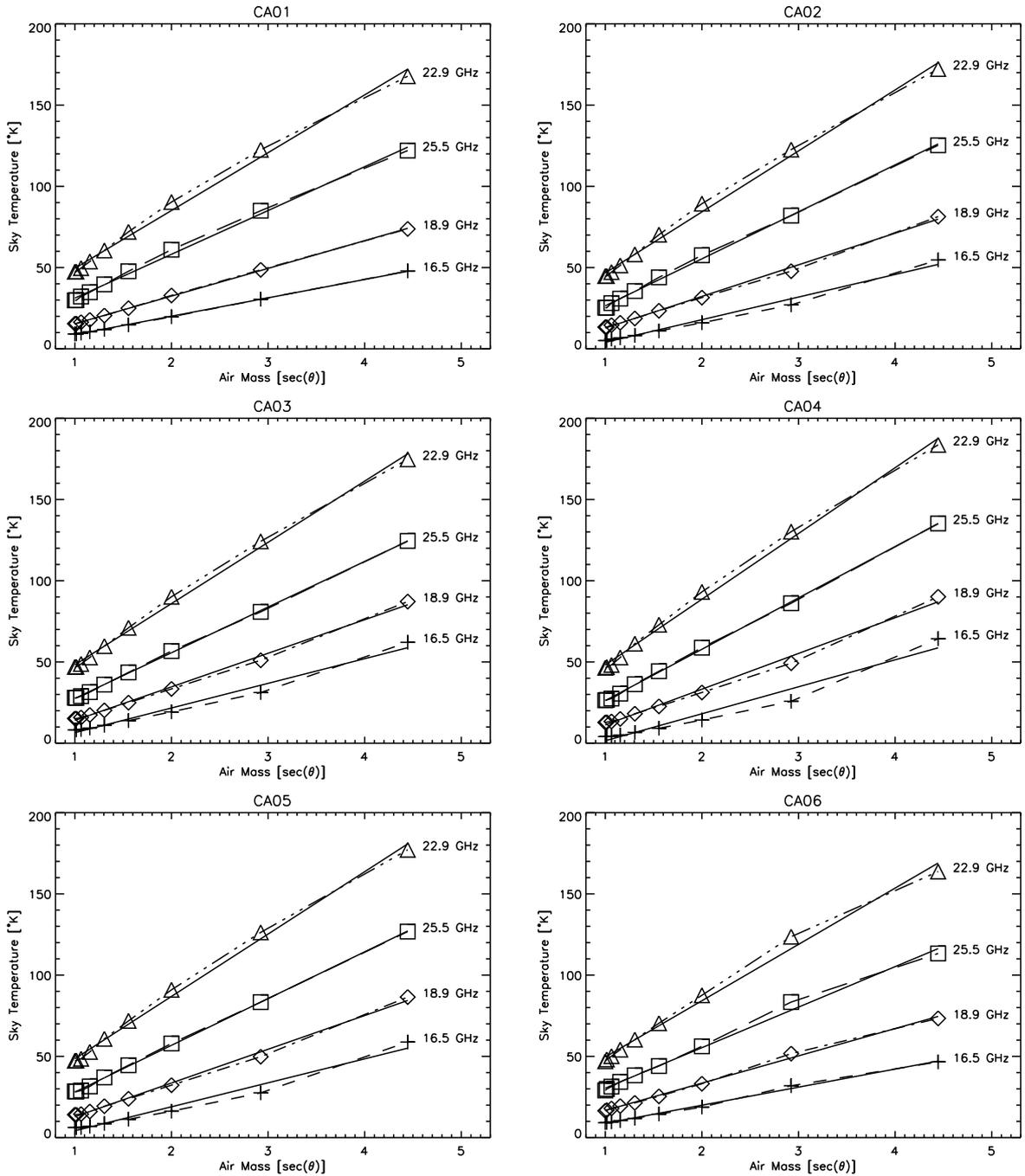}
\caption{Skydip functions measured for all the antennae on March 7 2011. The symbols (triangles, squares, diamond and crosses) show the data for each of the four filters (22.9, 25.5, 18.9, 16.5\,GHz, respectively), at each elevation (denoted by the airmass or sec($\theta$)).  The solid lines show the linear fits through the skydip data in each filter (i.e.\ eqn.~\ref{spillover2} and the dashed lines the fits to the full skydip function (eqn.~\ref{spillover}). The slopes yield the optical depths and the intersects with the $x$-axis (at sec($\theta$)=0) the spillover temperatures. The highest temperatures are measured at 22.9 GHz, in the middle of the water vapour line.}
\label{fig:skydip}
\end{figure}

\clearpage


\clearpage

\hspace{0.5cm}
\begin{figure}[h]
\centering
\includegraphics[width=1\columnwidth]{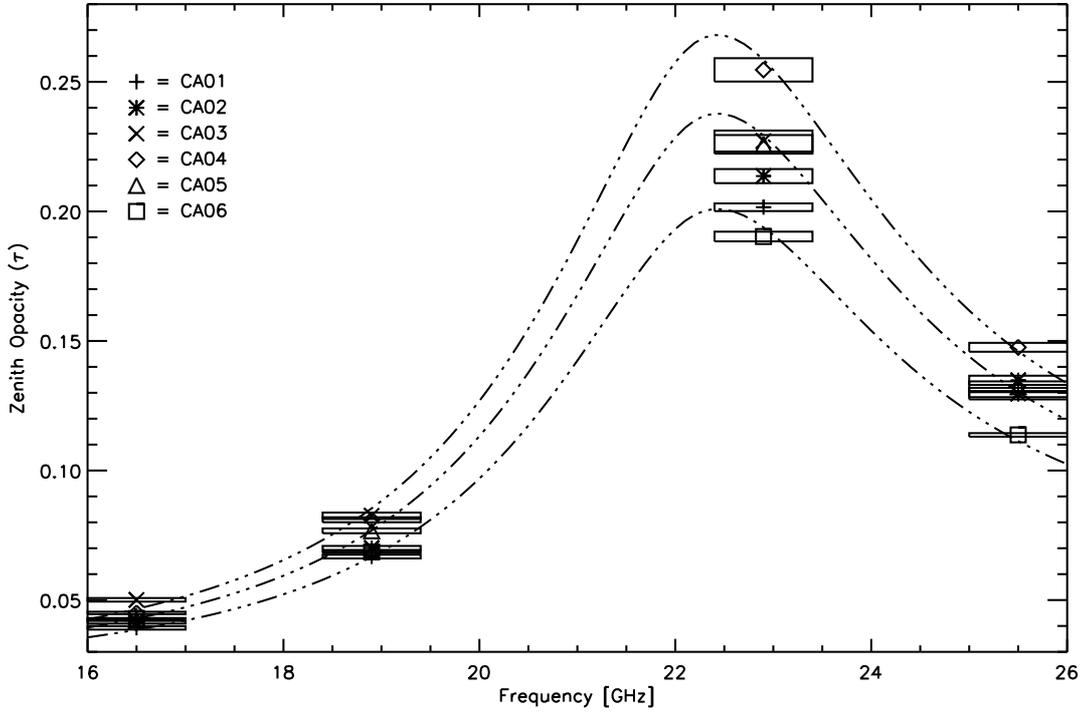}
\caption{Zenith opacities, $\tau$, as determined from the sky dips shown in Fig.~\ref{fig:skydip} and their values listed in Table \ref{tab:spillover}, compared to the opacties for three model atmospheres.  The opacities for each filter are shown as a box of 1\,GHz width and $\tau_{\mathsf{err}}$ height.  Additionally, the centre point is depicted with a symbol to allow identification of which antenna/WVR Unit the point belongs to, as indicated in the legend.  Over-plotted are the opacities derived from three ATM model atmospheres, for PWV values of 28\,mm (lowest opacity), 34\,mm (middle line) and 39\,mm (highest opacity). }
\label{fig:tau}
\end{figure}

\clearpage
\begin{figure}[h]
\centering
\includegraphics[width=1\columnwidth]{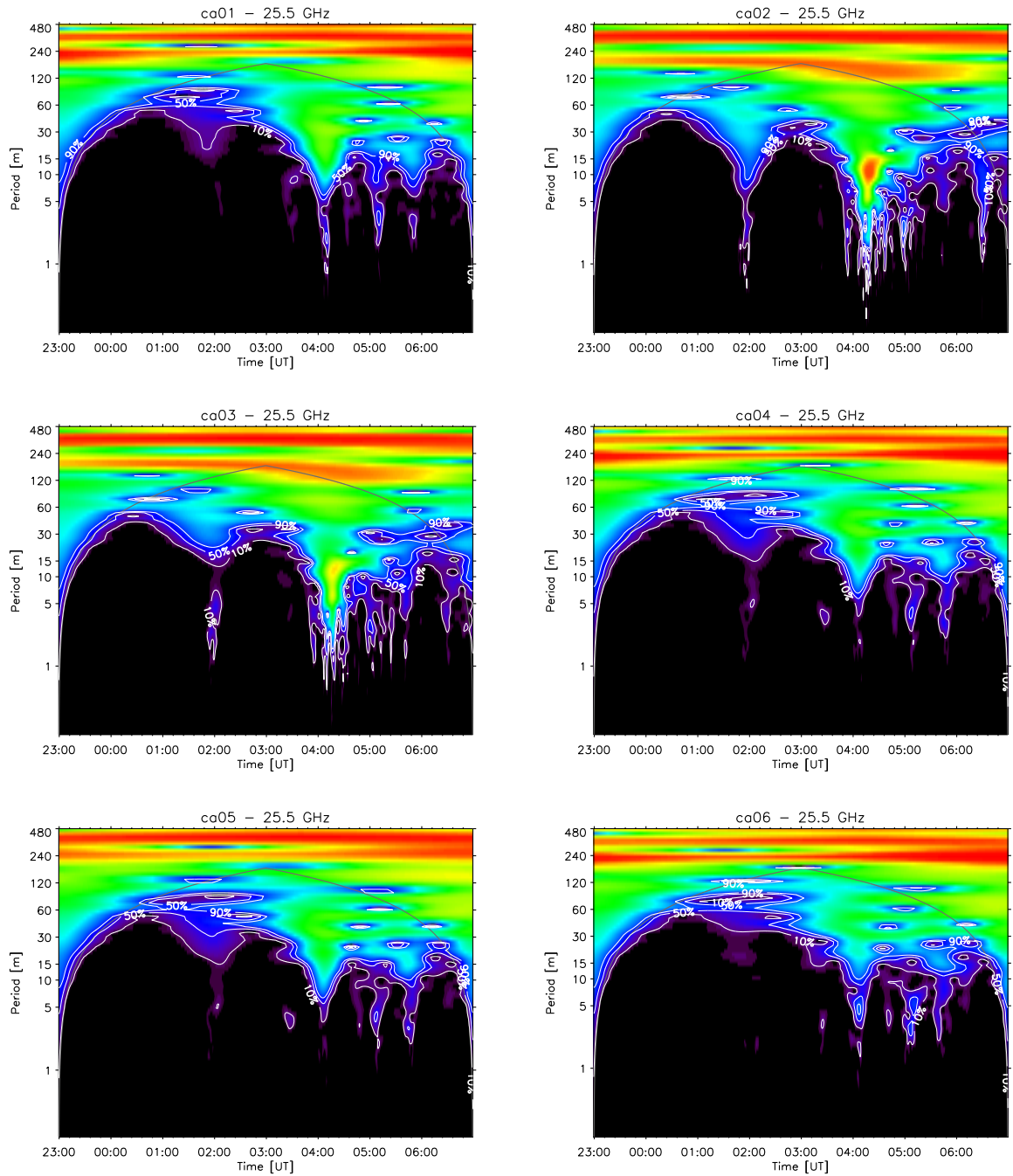}
\caption{The Morlet wavelet decomposition of a 12\,hr time series of raw voltages measured in the 25.5\,GHz channels on each antenna on 23 May 2012.  Antennae 2 and 3 were stowed, antennae 1, 4, 5 and 6 were tracking astronomical sources during this period. The $x$-axis is the time of day and the $y$-axis the period (in minutes) for the signal power. The solid lines marks the cone of influence arising from the edge effects, resulting from the finite size of the dataset. Above these lines, the results may be influenced by numerical artefacts, though their effects are clearly small here. The similarities of the transforms between the antennae pointing in the same directions are striking, as is the absence of any RFI. }
\label{wavelets_RF}
\end{figure}

\clearpage
\begin{figure}[h]
\centering
\includegraphics[width=1\columnwidth]{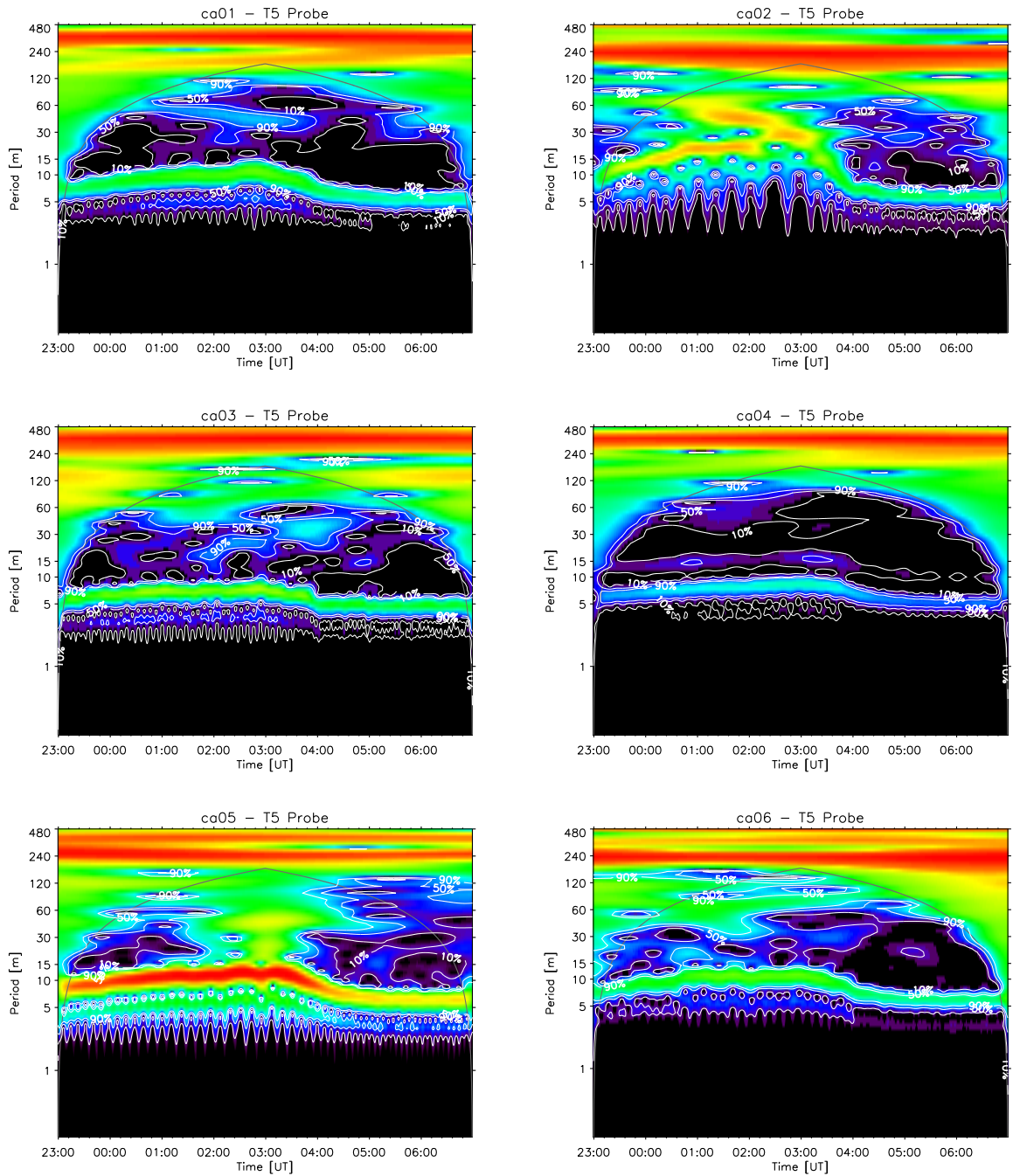}
\caption{The Morlet wavelet transform of the same time span as Figure \ref{wavelets_RF} but for the outer shell temperature of the WVR units (sensor T5). The activity and oscillatory behaviour of the air conditioning units in all antennae are clearly evident, as are the differences in the thermal behaviour between receiver cabins. This is in contrast to the sky brightness measurements, as shown in Fig.~\ref{wavelets_RF}.}
\label{wavelets_T5}
\end{figure}

\clearpage

\begin{figure}[h]
\centering
\includegraphics[width=1\columnwidth]{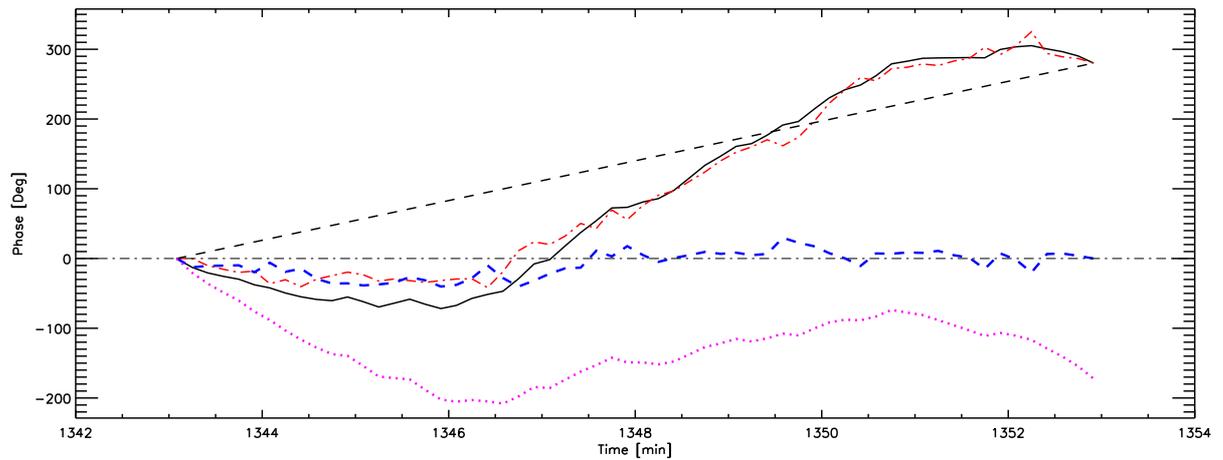}
\caption{Phase comparison of the strong calibrator 0537-441 at 48.3\,GHz on 3 June 2011 on a 4,500\,m long baseline between antennae 1 and 6.  
Shown are the calibrator phase -- solid (black) -- and the  WVR derived phase closely tracking the calibrator phase --  dash-dot (red).  The dashed line (blue, near phase angle $0^{\circ}$) quantifies the performance achieved by showing the calibrator phase minus WVR phase (i.e.\ the ``WVR residual phase''), while the calibrator phase minus the interpolated phase (i.e.\ the ``interpolated residual phase'') is shown using the dotted (magenta) line. This is obtained by subtracting the observed phase from the interpolated phase (dashed, black). This interpolated residual phase is the best result obtainable without the use of WVRs. The interpolated residual phase RMS is $47^{\circ}$, while the WVR residual phase RMS is $18^{\circ}$.   This corresponds to an improvement of over 40\% in correlation efficiency from $\epsilon=0.50$ uncorrected to $\epsilon=0.91$ after WVR correction.}
\label{fig:phase_plot_61}
\end{figure}

\clearpage

\begin{figure}[h]
\centering
\includegraphics[width=1\columnwidth]{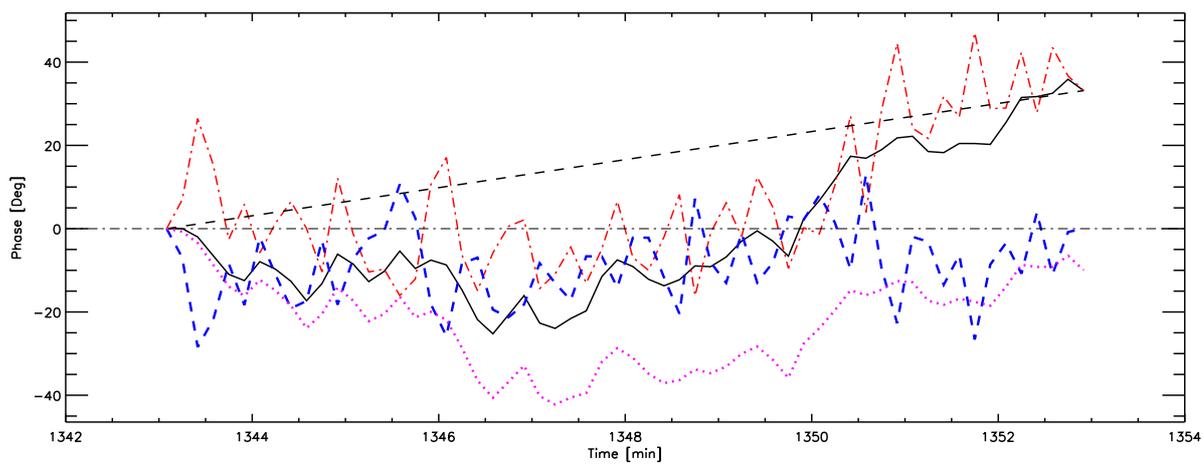}
\caption{Phase comparison of the strong calibrator 0537-441 at 48.3\,GHz on 3 June 2011 on a 92\,m short baseline between antennae 1 and 2.  The various lines shown are as for Fig.~\ref{fig:phase_plot_61}. The interpolated residual phase RMS is $11^{\circ}$, while the WVR residual phase RMS is $9^{\circ}$.   This corresponds to a negligible improvement of 1\% in correlation efficiency from $\epsilon=0.96$ uncorrected to $\epsilon=0.97$ after WVR correction.}
\label{fig:phase_plot_21}
\end{figure}

\clearpage

\begin{figure}[h]
\centering
\includegraphics[width=1\columnwidth]{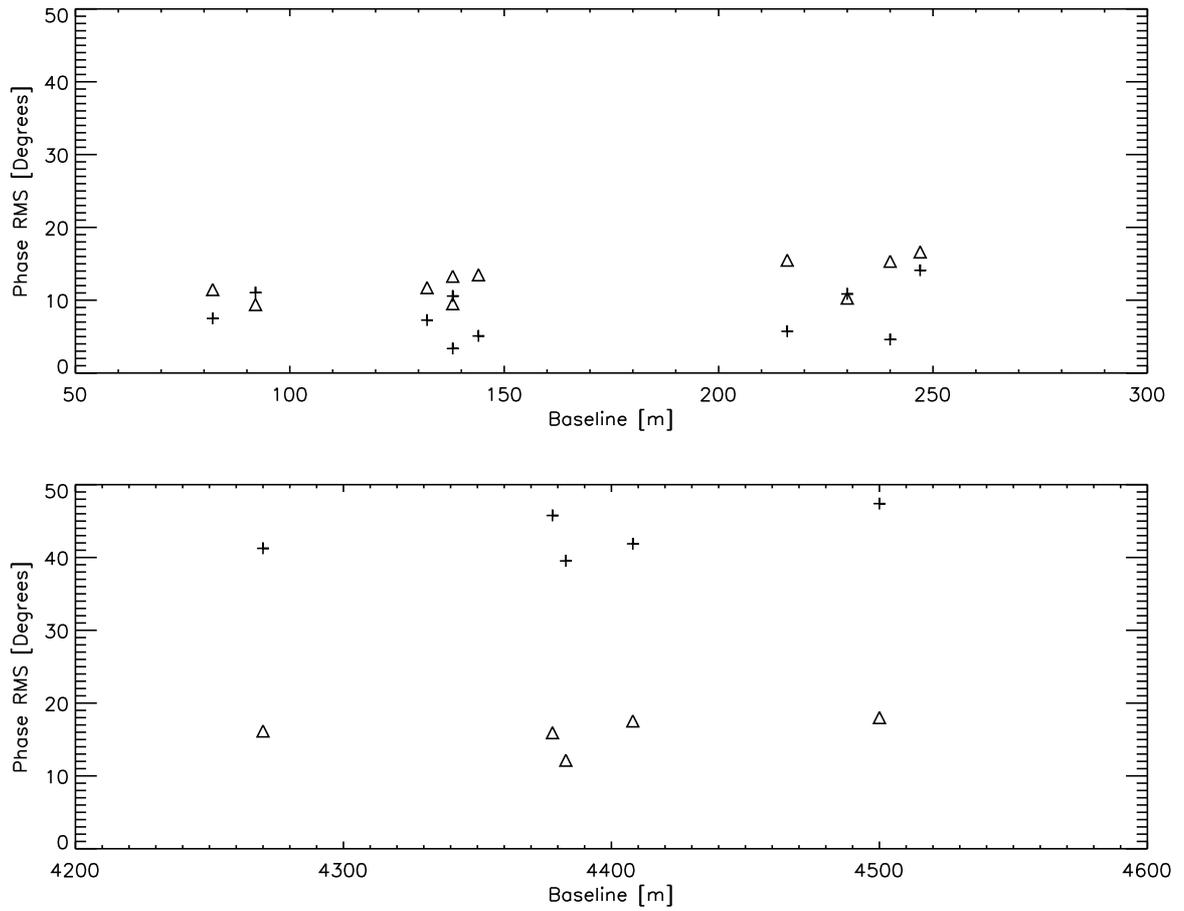}
\caption{The standard deviation of the WVR ($\triangle$'s) and interpolated ($+$'s) residual phases  for the data in Figs.~\ref{fig:phase_plot_61} and \ref{fig:phase_plot_21}, plotted against baseline length.  The upper plot is for the short baselines (up to 250\,m) and the lower plot is for the long baselines that use antenna 6 (i.e.\ $\sim 4$\,km). The clear gain achieved on the long baselines using the WVRs is readily apparent.}
\label{RMS_improvement}
\end{figure}

\clearpage